\documentclass[3p]{elsarticle}

\usepackage{amssymb}
\usepackage{amsmath}
\usepackage{amsfonts}
\usepackage{graphicx}
\usepackage{epstopdf}
\usepackage{subcaption}
\usepackage{hyperref}

\graphicspath{{../}}

\DeclareMathOperator*{\argmin}{\arg\!\min}

\journal{Applied and Computational Harmonic Analysis Letter}

\begin{document}

\begin{frontmatter}

\title{Parsimonious Representation of Nonlinear Dynamical Systems Through Manifold Learning: A Chemotaxis Case Study}

\author[princeton]{Carmeline~J.~Dsilva}
%
\author[technion]{Ronen~Talmon}
%
\author[yale]{Ronald R. Coifman}
%
\author[princeton,princetonpacm]{Ioannis~G.~Kevrekidis \corref{cor1}}
\ead{yannis@princeton.edu}

\address[princeton]{Department of Chemical and Biological Engineering, Princeton University, Princeton, NJ, 08540, USA}
\address[technion]{Department of Electrical Engineering, Technion - Israel Institute of Technology, Haifa, 3200003, Israel}
\address[yale]{Department of Mathematics, Yale University, New Haven, CT, 06520, USA}
\address[princetonpacm]{Program in Applied and Computational Mathematics, Princeton University, Princeton, NJ, 08540, USA}
\cortext[cor1]{Corresponding author}

\begin{abstract}
Nonlinear manifold learning algorithms, such as diffusion maps, have been fruitfully
applied in recent years to the analysis of large and complex data sets.
However, such algorithms still encounter challenges when faced with real data.
One such 
challenge is the existence of ``repeated eigendirections," which obscures the detection of the 
true dimensionality of the underlying manifold and arises when several embedding coordinates parametrize
the same direction in the intrinsic geometry of the data set. 
We propose an algorithm, based on local linear regression, 
to automatically detect coordinates corresponding to repeated eigendirections.
We construct a more parsimonious embedding using only the eigenvectors corresponding to unique eigendirections, and we show that this reduced diffusion maps embedding induces 
a metric which is equivalent to the standard diffusion distance.
We first demonstrate the utility and flexibility of our approach on synthetic data sets.
We then apply our algorithm to data collected from a stochastic model of cellular chemotaxis, where our approach for factoring out repeated eigendirections allows us to detect changes in dynamical behavior and the underlying
intrinsic system dimensionality directly from data.
\end{abstract}



\end{frontmatter}

\section{Introduction}

In recent years, data mining algorithms have proven useful for many disciplines and applications \cite{gepshtein2013image, fernandez2014diffusion, singer2011viewing, yuan2014automated, zhao2003face, trapnell2014dynamics, kemelmacher2011exploring, sifre2013rotation}.
For multiscale dynamical systems in particular, data-driven methodologies are essential for
assisting in model reduction when simple macroscopic models cannot be analytically obtained
due to the system complexity \cite{talmon2014intrinsic,berry2013time,singer2009detecting,ferguson2010systematic}.
For simulations or experimental observations of such systems, the detailed, microscale evolving system state is
very high-dimensional, and the reduction to useful macroscale dynamics is not obvious.
In such cases, data obtained from observations and/or simulations of the dynamical system combined with
data mining methodologies can lead to a low-dimensional description which not only provides insight into the underlying dynamics, but also serves as a first step in constructing macroscale models consistent with the observed microscale behavior.

%
Several manifold learning algorithms obtain parametrizations of the data through
the spectral analysis of a Laplace operator \cite{Belkin2003,coifman2005geometric,coifman2006geometric,singer2008non}.
The data is then embedded in a new low-dimensional coordinate system
given by the eigenvectors of this operator.
The premise is that these coordinates, obtained in a data-driven manner, are the right macroscopic ``observables", i.e., 
the variables which parametrize the macroscale dynamical behavior of the system, thus enabling the construction of a reduced macroscale model.

Such manifold learning algorithms were initially applied to synthetic data sets, to illustrate
their geometric properties and flexibility \cite{coifman2005geometric, nadler2006diffusion}.
More recently they have been applied to experimental as well as simulation data,
enabled by advances in data representation (observers) and
metrics \cite{rubner2000earth,mallat2012group,talmon2013empirical,zhao2014rotationally, rohrdanz2011determination, talmon2015manifold}.
A nontrivial shortcoming of these methods is that, from a geometric perspective,
not all eigenvectors are guaranteed to parametrize unique directions within the data;
some eigenvectors are ``repeated eigendirections'' which describe the same coordinate in the intrinsic geometry
of the data.
Identifying these eigenvectors is critical for obtaining a
parametrization of the system which captures the true dimensionality of the macroscale dynamics.
This task is often done manually, and few methods have been proposed to automate the identification of the unique eigendirections \cite{gerber2007robust}.

In this paper, we propose an algorithm to automatically identify such unique eigendirections using local linear regression \cite{wasserman2006all}.
We first demonstrate and validate our algorithm on synthetic examples, where a closed-form solution for the
eigenvectors and eigenvalues of the Laplace operator is known.
We then consider a simulation data set from a stochastic dynamical system modeling
cellular chemotaxis \cite{othmer1988models}.
Recent advances in observers and metrics, coupled with the proposed approach for identifying the unique eigendirections,
provide a data analysis pipeline which successfully analyzes this simulation in a purely data-driven manner.
We will show that this pipeline allows us to detect changes in the dimensionality of the underlying macroscale dynamics, which are
related to changes in the regime/mode of the system.

\section{Manifold Learning Based on Laplace Operators}
%
%

Let $z(1), \dots, z(m) \in \mathbb{R}^n$ denote $m$ state observations sampled from an evolving autonomous dynamical system.
We assume the $n$-dimensional observations $z(i)$ lie on a $d$-dimensional manifold $\mathcal{M}_d$, where $d < n$.
We therefore consider the problem of parametrizing the continuous $d$-dimensional manifold $\mathcal{M}_d$ embedded in $\mathbb{R}^n$ from data.
For linear hyperplanes, the principal axes parametrize the manifold.
However, in the case when the manifold is nonlinear, a set of appropriate coordinates is not readily apparent.
By using a particular manifold learning technique based on the construction of a Laplace operator, namely, {\em diffusion maps}, we will show how to extract a $d$-dimensional parametrization of the observations which is consistent with the geometry of the manifold \cite{Belkin2003, coifman2005geometric, singer2008non}.

\subsection{Eigenfunctions of the Laplace-Beltrami operator}

\begin{figure}[t]
\centering
\begin{subfigure}{0.3\textwidth}
\includegraphics[width=\textwidth]{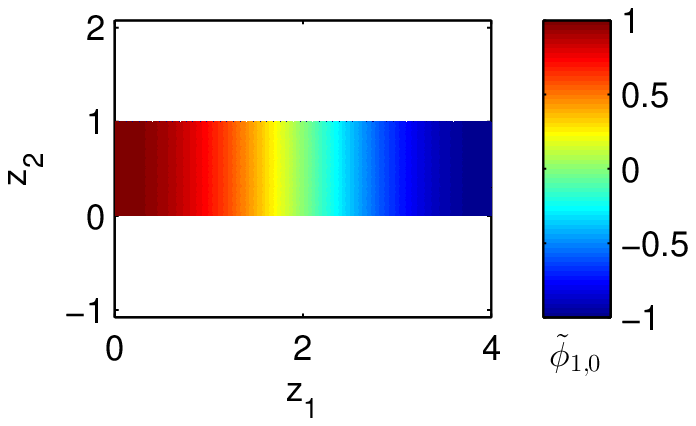}
\includegraphics[width=\textwidth]{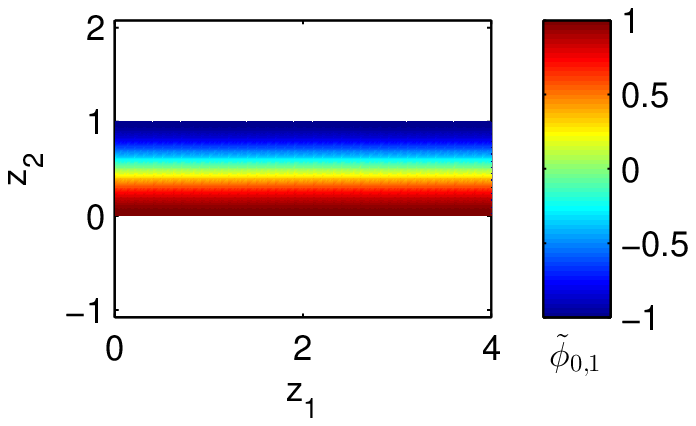}
\caption{}
\label{subfig:strip_efuncs}
\end{subfigure}
\begin{subfigure}{0.3\textwidth}
\includegraphics[width=\textwidth]{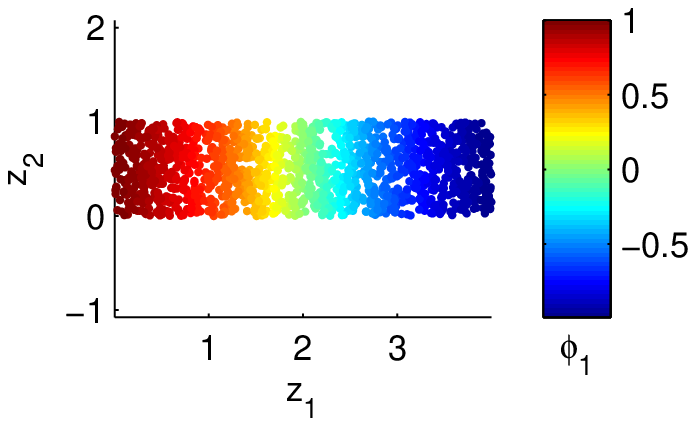}
\includegraphics[width=\textwidth]{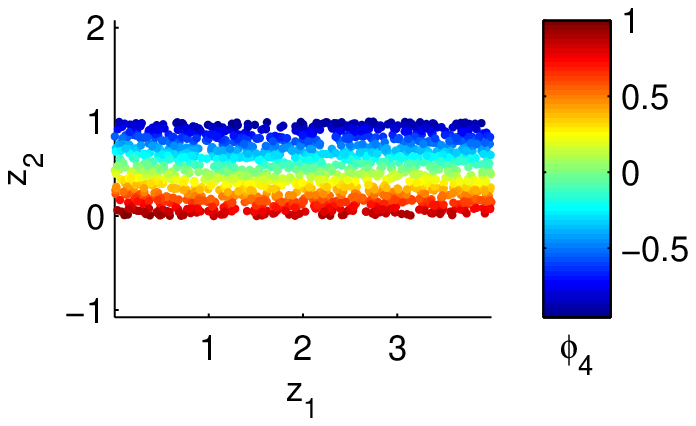}
\caption{}
\label{subfig:strip_evecs_uniform}
\end{subfigure}
\begin{subfigure}{0.3\textwidth}
\includegraphics[width=\textwidth]{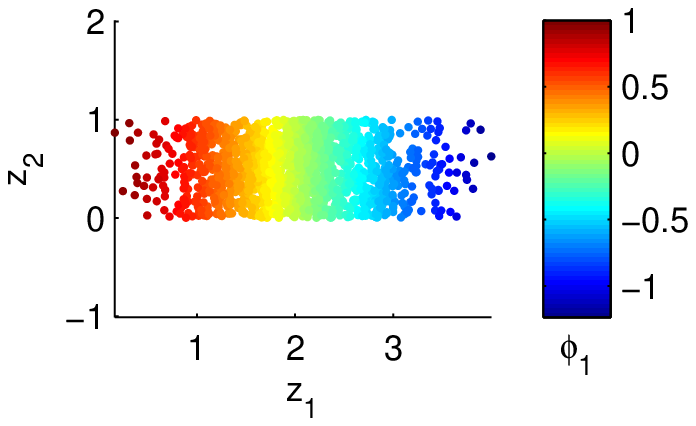}
\includegraphics[width=\textwidth]{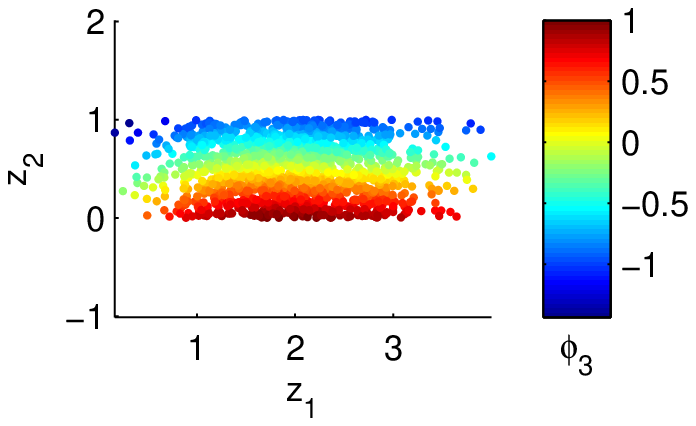}
\caption{}
\label{subfig:strip_evecs_nonuniform}
\end{subfigure}
\caption{(a) Two-dimensional continuous strip colored by the eigenfunctions $\tilde{\phi}_{1, 0} = \cos \left( {\pi z_1}/{L_1} \right)$, and $\tilde{\phi}_{0, 1} = \cos \left( {\pi z_2}/{L_2} \right)$. (b) Data, uniformly sampled from a two-dimensional strip, colored by the first and fourth (non-trivial) eigenvectors of the discrete Laplacian. (c) Data, sampled from a Gaussian distribution in $z_1$ and sampled uniformly in $z_2$, colored by the first and third (non-trivial) eigenvectors of the discrete Laplacian. Note that in all cases we uncover parametrizations which are one-to-one with $z_1$ and $z_2$ on the domain.}
\end{figure}

In recent years it has been noted that the eigenfunctions of the \emph{continuous}
Laplace-Beltrami operator provide ``good" coordinates for a manifold \cite{jones2008}.
We start by considering such a continuous setting, where rigorous analysis is possible for specific examples.
Using such an example, we will demonstrate that the existence of repeated eigendirections is
inherent to the manifold learning setup based on Laplace operators,
and that the identification of the unique eigendirections requires some additional effort.
The fact that these challenges arise in the continuous setting implies that, even in the limit of infinite data, such repeated eigendirections still pose a problem for data analysis.

To review and illustrate why these eigenfunctions provide appropriate coordinates, consider a two-dimensional strip with edge lengths $L_1$ and $L_2$.
The eigenvalues of the Laplace-Beltrami operator with Neumann boundary conditions are given by
\begin{equation} \label{eq:evals}
\tilde{\mu}_{k_1, k_2} = \left( \frac{k_1 \pi}{L_1} \right)^2 + \left( \frac{k_2 \pi}{L_2} \right)^2
\end{equation}
for $k_1, k_2 = 0, 1, 2, \dots$,
and the corresponding eigenfunctions are
\begin{equation} \label{eq:efuncs}
\tilde{\phi}_{k_1, k_2} = \cos \left( \frac{k_1 \pi z_1}{L_1} \right) \cos \left( \frac{k_2 \pi z_2}{L_2} \right)
\end{equation}
where $z_1$ and $z_2$ denote the two coordinates of the strip \cite{singer2008non}.
We note that the eigenfunctions $\tilde{\phi}_{1, 0} = \cos \left( {\pi z_1}/{L_1} \right)$ and
$\tilde{\phi}_{0, 1} = \cos \left( {\pi z_2}/{L_2} \right)$ are one-to-one with the $z_1$ and $z_2$ coordinates,
respectively, on the domain, and therefore provide a parametrization of the underlying manifold (see Figure~\ref{subfig:strip_efuncs}).
Furthermore, the corresponding eigenvalues $\tilde{\mu}_{1,0}$ and $\tilde{\mu}_{0,1}$ provide an estimate of the
relative magnitude of $L_1$ versus $L_2$: as the ratio between $L_1$ and $L_2$ increases,
the gap between $\tilde{\mu}_{1,0}$ and $\tilde{\mu}_{0,1}$ also increases
(this will be discussed further in Section~\ref{sec:relative_lengths}).
The analytic form of the eigenfunctions in \eqref{eq:efuncs} illustrates the two issues we address in this paper.
First, $z_1$ and $z_2$ are not necessarily decoupled in subsequent eigenfunctions,
and a proper parametrization of the manifold is not necessarily given by the $d$ eigenfunctions associated with the
smallest $d$ eigenvalues.
%
Second, eigenfunctions with $k_1+k_2 \ge 2$ do not describe any
additional directions intrinsic to the strip geometry; we will refer to these as ``repeated eigendirections,''
and we will refer to the eigenfunctions with $k_1+k_2 =1$ as ``unique eigendirections.''
Clearly, this problem of repeated eigendirections also arises for more curved of more than two variables.
We note that, although the eigenfunctions can only be written analytically for very special cases,
it has been observed empirically that they often provide appropriate coordinates
to usefully parametrize more complicated, nonlinear manifolds.

\subsection{Discrete approximation of the Laplace-Beltrami operator: diffusion maps}

In most applications, we are not given a description of the continuous manifold on which the data lie.
Instead, we are given data {\em sampled} from the underlying manifold,
and the parametrization of the manifold needs to be uncovered from the data.
This can be accomplished by constructing a matrix which approximates the Laplace-Beltrami operator.
It was shown in \cite{coifman2006geometric} that, in the limit of infinite data,
this discrete Laplacian matrix constructed from data converges pointwise to the continuous Laplace-Beltrami operator on the manifold.
As a result, the eigenvectors of the discrete Laplacian approximate the eigenfunctions of this continuous operator.

Given observations $z(1), \dots, z(m) \in \mathcal{M}_d$, we first construct the weight matrix $\mathbf{W} \in \mathbb{R}^{m \times m}$, with
\begin{equation} \label{eq:W}
\mathbf{W}_{ij} = \exp \left( -\frac{\|z(i) - z(j) \|^2}{\epsilon^2} \right), \ i,j=1,\ldots,m,
\end{equation}
where $\| \cdot \|$ denotes the appropriate norm for the observations, and $\epsilon$ is a characteristic distance between the observations.
The kernel's scale $\epsilon$ can be chosen using several heuristics \cite{rohrdanz2011determination, coifman2008graph};
we often take $\epsilon$ to be the median of the pairwise distances between the data points. 
The underlying assumption is that, even in cases where the samples embedded in the high dimensional space lie on a highly nonlinear manifold, within $\epsilon$-local neighborhoods, the Euclidean distance respects the tangent plane to the manifold and thus locally conveys meaningful relationships,. 
We then construct the diagonal matrix $\mathbf{D} \in \mathbb{R}^{m \times m}$, with $\mathbf{D}_{ii} = \sum_j \mathbf{W}_{ij}$,
and form the matrix $\widetilde{\mathbf{W}} = \mathbf{D}^{-\alpha} \mathbf{W} \mathbf{D}^{-\alpha}$, where $0 \le \alpha \le 1$.
Next, we construct the diagonal matrix $\widetilde{\mathbf{D}} \in \mathbb{R}^{m \times m}$, with $\widetilde{\mathbf{D}}_{ii} = \sum_j \widetilde{\mathbf{W}}_{ij}$, and the matrix $\mathbf{A}  = \widetilde{\mathbf{D}}^{-1} \widetilde{\mathbf{W}}.$

If the data $z(1), \dots, z(m)$ are sampled from $\mathcal{M}_d$ with some density $q$, then, for $\epsilon \rightarrow 0$ and $m \rightarrow \infty$ (with the appropriate rates), the discrete matrix converges to the following continuous limit operators with Neumann boundary conditions (as discussed in the previous section) \cite{coifman2006geometric}
\begin{align} 
\label{eq:limiting_op1}
\frac{1}{\epsilon^2}(\mathbf{I}-\mathbf{A}) \phi &\rightarrow \nabla^2 \phi - 2\nabla U \cdot \nabla \phi, &&\alpha = 0 \\
\label{eq:limiting_op2}
\frac{1}{\epsilon^2}(\mathbf{I}-\mathbf{A}) \phi &\rightarrow \nabla^2 \phi - \nabla U \cdot \nabla \phi, &&\alpha = 1/2 \\
\label{eq:limiting_op3}
\frac{1}{\epsilon^2}(\mathbf{I}-\mathbf{A}) \phi &\rightarrow \nabla^2 \phi, &&\alpha = 1
\end{align}
where $U = - \log q$.
The different limit operators, depending on the choice of $\alpha$, imply that nonuniform
sampling on the manifold may have different effects via the potential $U$ in \eqref{eq:limiting_op1}--\eqref{eq:limiting_op3}.
In particular, by setting $\alpha=1$, one can factor out the density effects in the
weight matrix, and the discrete Laplacian matrix approaches the Laplace-Beltrami operator on the manifold $\mathcal{M}_d$.
Therefore, the eigenvectors $\phi_0, \phi_1, \dots, \phi_{m-1}$ of $\mathbf{A}$ approximate the eigenfunctions of the Laplace-Beltrami operator on $\mathcal{M}_d$,
and the eigenvalues $\mu_0, \mu_1, \dots, \mu_{m-1}$ of $\mathbf{A}$ are related to the eigenvalues of the continuous operator by
\begin{equation} \label{eq:evals_relationship}
\mu_k = \exp \left( -\frac{\epsilon^2}{4} \tilde{\mu}_{k_1, k_2}  \right).
\end{equation}

As discussed previously, the eigenfunctions provide a parametrization of the manifold,
such that $\mu_j^\tau \phi_{j}(i)$ yields the $j^{th}$ embedding coordinate for $z(i)$, where $\tau \in \mathbb{N}$ is
a parameter (in our examples, we will take $\tau=0$).
The parameter $\tau$ corresponds to the number of iterates of the Laplace operator, or in other words, the number of diffusion steps (for more details, see \cite{coifman2006geometric}). 
%
%
%
This mapping $z(i) \mapsto \left(\mu_1^\tau \phi_{1}(i), \dots, \mu_{m-1}^\tau \phi_{m-1}(i)\right)$ is the standard diffusion maps embedding \cite{coifman2005geometric, coifman2006geometric}, where
we order the eigenvectors such that $|\mu_0| \ge |\mu_1| \ge \dots \ge |\mu_{m-1}|$.
Because the matrix $\mathbf{A}$ is row-stochastic ($\sum_j \mathbf{A}_{ij} = 1$),  $\mu_0 = 1$ and $\phi_0$ is a trivial constant vector.
The distance induced by this mapping is called the standard diffusion distance,
\begin{equation}
D^2_\tau(z(i), z(j)) = \sum_{k=1}^{m-1} \mu_k^{2 \tau} \left( \phi_k(i) - \phi_k(j)  \right)^2.
\end{equation}
Figures~\ref{subfig:strip_evecs_uniform}--\ref{subfig:strip_evecs_nonuniform} shows
data sampled from a strip, colored by eigenvectors of $\mathbf{A}$.
In cases of both uniform and nonuniform sampling, the selected eigenvectors are one-to-one with $z_1$ and $z_2$,
and thus parametrize the manifold.
%
%
Although we have considered a very simple example for illustrative purposes, common practice is to use these tools for high-dimensional, nonlinear data sets.

From the previous section, we know that the eigenfunctions with $(k_1, k_2) =(1, 0)$ and $(k_1, k_2) =(0, 1)$
provide embedding coordinates for the manifold; these two eigenfunctions are both uncoupled and not repeated.
From \eqref{eq:evals}, we see that sorting the eigenvectors by the magnitude of
the corresponding eigenvalues implies that these two eigenvectors are guaranteed to
appear before any coupled or repeated eigendirections.
However, these eigenvectors are {\em not} guaranteed to appear as {\em the first two} (non-trivial) eigenvectors,
as harmonics of the first eigendirection (i.e., $\cos \left( n \pi z_1 / L_1 \right)$ with $n > 1$) could appear before the second.
We note that, in contrast to our simple illustrative example, for most data sets of interest, the coordinates for the underlying manifold are unknown and cannot easily be obtained from the coordinates of the original data, so that identifying which eigenvectors correspond to unique eigendirections is nontrivial.

\section{Identifying the Informative Eigenvectors }

\begin{figure}[t]
\centering
\begin{subfigure}{0.24\textwidth}
\includegraphics[width=\textwidth]{strip_discrete1}
\end{subfigure}
\begin{subfigure}{0.24\textwidth}
\includegraphics[width=\textwidth]{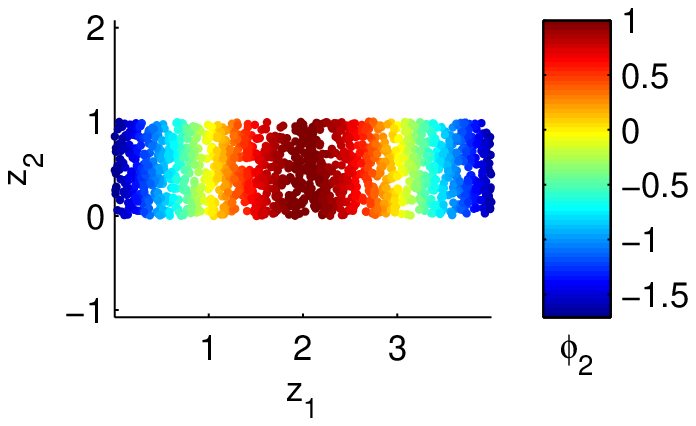}
\end{subfigure}
\begin{subfigure}{0.24\textwidth}
\includegraphics[width=\textwidth]{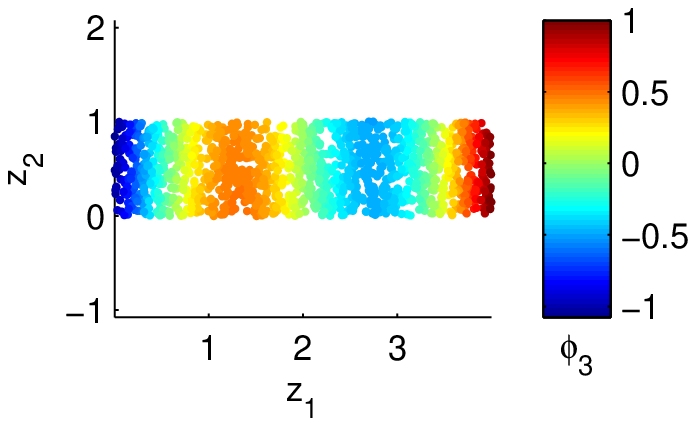}
\end{subfigure}
\begin{subfigure}{0.24\textwidth}
\includegraphics[width=\textwidth]{strip_discrete4}
\end{subfigure}
\caption{ Data, uniformly sampled from a two-dimensional strip, colored by the first four (non-trivial) eigenvectors from diffusion maps. Note that the first and fourth eigenvectors are one-to-one with $z_1$ and $z_2$, respectively. However, the second and third eigenvectors are higher harmonics of the first eigenvector and do not capture any additional structure within the data set. }
\label{fig:strip_harmonics}
\end{figure}

Common practice is to order the eigenvectors by the magnitude of the corresponding eigenvalues,
assuming that the leading eigenvectors provide a parametrization of the underlying manifold.
However, as discussed in the previous section, some eigenvectors are higher harmonics of
previous eigenvectors and do not describe new directions in the data set \cite{gerber2007robust}.
For the case where $\mathcal{M}_d$ is a $2$-dimensional strip with edge lengths $L_1  > L_2$, recall
that the eigenfunctions $\tilde{\phi}_{1,0} = \cos \left(  {\pi z_1}/{L_1} \right)$ and
$\tilde{\phi}_{0,1} = \cos \left(  {\pi z_2}/{L_2} \right)$ provide embedding coordinates for the manifold $\mathcal{M}_d$.
However, these two eigenvectors $\tilde{\phi}_{1, 0}$ and $\tilde{\phi}_{0, 1}$ are not guaranteed to
correspond to the two smallest (non-trivial) eigenvalues (the smallest eigenvalue will always be $\tilde{\mu}_{0,0} = 0$ and
correspond to a constant eigenfunction; these eigenvalues are related to the eigenvalues $\mu_k$ of the discrete
operator via \eqref{eq:evals_relationship}).
In fact, if $L_1 > 2 L_2$, then $\tilde{\mu}_{2, 0} < \tilde{\mu}_{0, 1}$, and so the
second (non-trivial) eigenvector (when the eigenvectors are ordered by their corresponding eigenvalues)
will be a ``repeated eigendirection" of the first and still parametrize $z_1$ (see Figure~\ref{fig:strip_harmonics}).
Automatic detection of which eigenvectors are harmonics of each other is clearly useful.
Utilizing only the eigenvectors $\phi_i$ which correspond to unique eigendirections yields
the most parsimonious representation of the data.
We will show how we can automatically detect these eigenvectors to obtain a meaningful representation of the data,
and that using only these eigenvectors yields a reduced embedding which is equivalent to the standard diffusion maps embedding.
Furthermore, when the data is uniformly sampled from the underlying manifold, the corresponding eigenvalues provide us with an estimate of the relative lengths of the data set along these unique eigendirections.

\subsection{Algorithm: local linear regression}

Given the eigenvectors $\phi_1, \phi_2, \dots, \phi_{m-1} \in \mathbb{R}^m$, we would like to automatically deduce
which ones capture new directions in the data, and which ones are merely repeated eigendirections.
This problem was addressed previously in \cite{gerber2007robust} by performing
successive iterations of diffusion maps, interspersed with advection along the first eigendirection at each iteration.
However, both the advection
procedure and the successive eigendecompositions are expensive and soon become intractable for larger data sets.
Here, we propose an alternative approach to address the problem of repeated eigendirections
motivated by simple trigonometric arguments (for example, the repeated eigendirection $\cos2x$ 
can be written as a function of the unique eigendirection $\cos x$).
We therefore attempt to fit a function $f(\phi_1, \dots, \phi_{k-1})$ to $\phi_{k}$; if the
resulting fit is accurate, we assume $\phi_{k}$ is a repeated eigendirection of $\phi_1, \dots, \phi_{k-1}$.
We use a {\em local} linear function
\begin{equation}
\phi_k(i) \approx \alpha_k(i) + \beta_k^T(i) \Phi_{k-1}(i)
\end{equation}
as our functional approximation, where
$\Phi_{k-1}(i) = \begin{bmatrix} \phi_1(i) & \dots & \phi_{k-1}(i) \end{bmatrix}^T$,
$\alpha_k(i) \in \mathbb{R}$, and $\beta_k(i) \in \mathbb{R}^{k-1}$.
The coefficients $\alpha_k(i)$ and $\beta_k(i)$ are not constant because we use a {\em local} linear fit in the $k-1$-dimensional $\Phi_{k-1}$ space, and so the coefficients change as a function of the domain.

At each point $\Phi_{k-1}(i)$, we approximate $\phi_k(i)$ by fitting a local linear function using the remaining $m-1$ data points.
We solve the following optimization problem
\begin{equation} \label{eq:opt_problem}
\hat{\alpha}_k (i) , \hat{\beta}_k(i)  = \argmin_{\alpha, \beta} \sum_{j \ne i} K(\Phi_{k-1}(i), \Phi_{k-1}(j)) \left( \phi_{k}(j) - (\alpha + \beta^T \Phi_{k-1}(j)) \right)^2.
\end{equation}
where $K$ is a kernel weighting function.
We use a Gaussian kernel,
\begin{equation}
K(\Phi_{k-1}(i), \Phi_{k-1}(j))  = \exp \left( - \frac{\|\Phi_{k-1}(i) - \Phi_{k-1} (j) \|^2}{\epsilon_{reg}^2} \right),
\end{equation}
where $\epsilon_{reg}$ is the kernel scale for the regression algorithm.
We typically take $\epsilon_{reg} = M / 3$, where $M$ is the median of the pairwise distances between $\Phi_{k-1}(i)$,
as we empirically found this choice to yield good results.
We then define the normalized leave-one-out cross-validation error for this local linear fit as
\begin{equation} \label{eq:cv_error}
r_{k} = \sqrt{ \frac{\sum_{i=1}^n \left( \phi_{k} (i) - (\hat{\alpha}_k(i) + \hat{\beta}_k(i)^T \Phi_{k-1}(i))  \right)^2} {\sum_{i=1}^n  \left( \phi_{k} (i) \right)^2 }}.
\end{equation}
Note that a small value of $r_k$ implies that $\phi_{k}$ can be accurately approximated from $\phi_1, \dots, \phi_{k-1}$.
We reason that a small value of $r_k$ implies that $\phi_k$ is a harmonic of previous modes, i.e.,
a repeated eigendirection, and conversely, a large value of $r_{k}$ indicates that $\phi_{k}$
parametrizes a new unique eigendirection in the data.
We choose to set $r_1 = 1$.
The error in \eqref{eq:cv_error} can easily be computed (see Section~5.4 of \cite{wasserman2006all}; code is also available at \url{http://ronen.eew.technion.ac.il/sample-page/journal/}).


We propose using {\em only} those eigenvectors for which $r_k$ is large to embed the data,
as this will yield a more parsimonious representation.
However, it is important to note that such an embedding also preserves certain features of the standard diffusion map embedding.
Let $I = \{i_1, i_2, \dots, i_d \}$ denote the indices of the identified unique eigendirections (i.e., $r_{i_1}, \dots, r_{i_d}$ are large).
Under the assumption that $\phi_j = f \left( \phi_{i_1}, \dots, \phi_{i_d} \right)$ for $j \not\in I$, where $f$ is Lipschitz continuous with Lipschitz constant $K$, one can show that
\begin{equation}
\frac{1}{1+K^2 \sum_{k \not\in I} \mu_k^{2\tau}} D^2_\tau(z(i), z(j)) \le \sum_{k \in I} \mu_k^{2 \tau} \left( \phi_k(i) - \phi_k(j)  \right)^2 \le D^2_\tau(z(i), z(j)).
\end{equation}
Therefore (for finite data, or provided the eigenvalues $\mu_k$ decay sufficiently fast in the limit of infinite data),
the distance induced by the $d$ eigenvectors $\phi_{i_1}, \dots, \phi_{i_d}$ identified as
parametrizing unique eigendirections is equivalent to the standard diffusion distance.
We will refer to this distance
\begin{equation}
\tilde{D}^2_\tau(z(i), z(j)) = \sum_{k \in I} \mu_k^{2 \tau} \left( \phi_k(i) - \phi_k(j)  \right)^2
\end{equation}
as the {\em reduced} diffusion distance,
and the embedding obtained from the corresponding eigenvectors as the {\em reduced} diffusion maps embedding.

\subsection{Estimating the ``relative lengths''} \label{sec:relative_lengths}

For a two-dimensional strip, once the eigenvectors which parametrize unique eigendirections have been determined,
the corresponding eigenvalues can be used to calculate the relative lengths along these directions.
For more general manifolds (at least for the case of uniform sampling), we postulate that the eigenvalues can still provide a measure of the relative
significance of the unique eigendirections.
Let $\mu_{i_1}, \mu_{i_2}, \dots$ denote the eigenvalues corresponding to
eigenvectors which parametrize unique eigendirections (i.e., those eigenvectors where $r_{i_j}$ is large).
For the strip example, from \eqref{eq:evals} and \eqref{eq:evals_relationship}, we propose to approximate the relative lengths $L_j$  along the manifold by
\begin{equation} \label{eq:est_lengths}
L_j \propto \frac{1}{\sqrt{-\log \mu_{i_j}}}.
\end{equation}
These lengths can then be used to evaluate how many components are required to retain
most of the information within the data set, similar to the eigenvalues in principal component analysis.

\subsection{Illustrative examples} \label{sec:illustrative_examples}

We demonstrate our proposed approach on three synthetic data sets.
The first data set is the two-dimensional strip discussed previously, where the eigenvalues and eigenvectors are known analytically.
The second and third data sets involve nonlinear manifolds which demonstrate the flexibility of our approach.

\subsubsection{Strip}

\begin{figure}[!t]
\centering
\begin{subfigure}{0.25\textwidth}
\includegraphics[height=2.5cm]{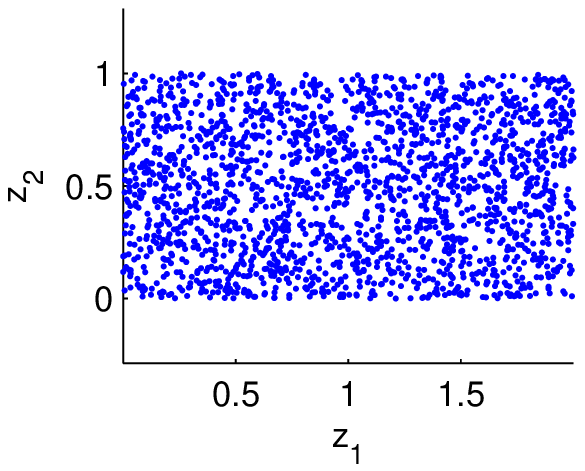}
\includegraphics[height=3cm]{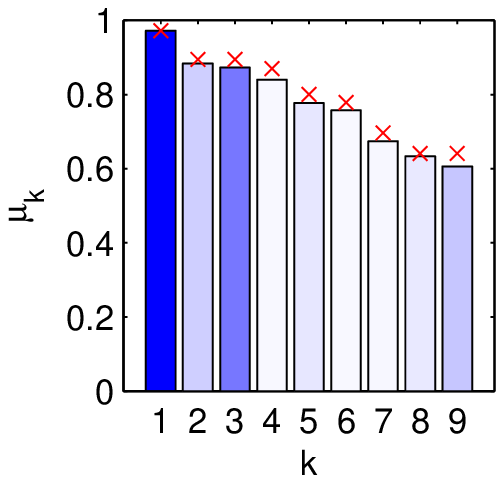}
\caption{}
\end{subfigure}
%
%
\begin{subfigure}{0.25\textwidth}
\includegraphics[height=2.5cm]{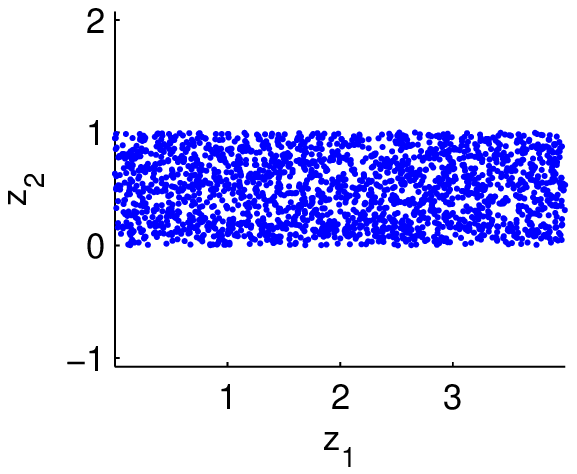}
\includegraphics[height=3cm]{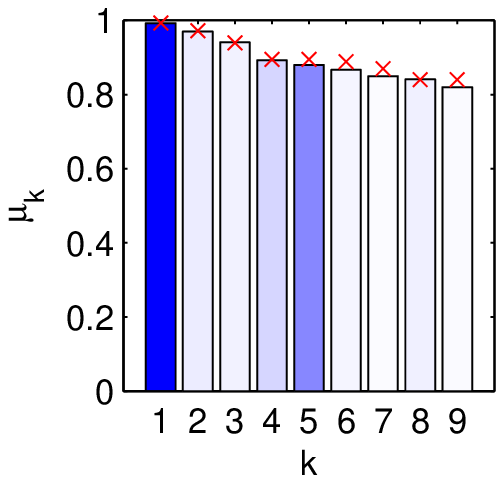}
\caption{}
\end{subfigure}
%
%
\begin{subfigure}{0.3\textwidth}
\includegraphics[height=2.5cm]{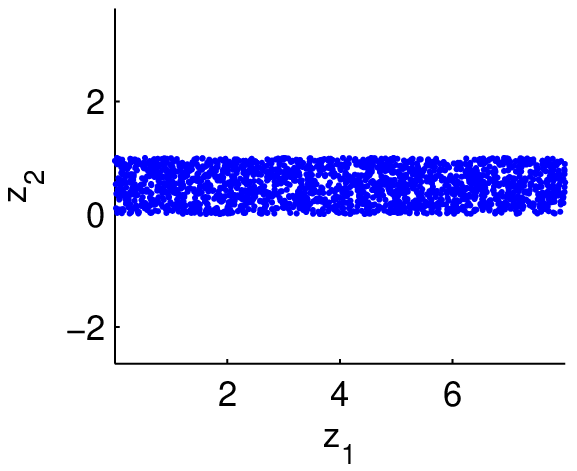}
\includegraphics[height=3cm]{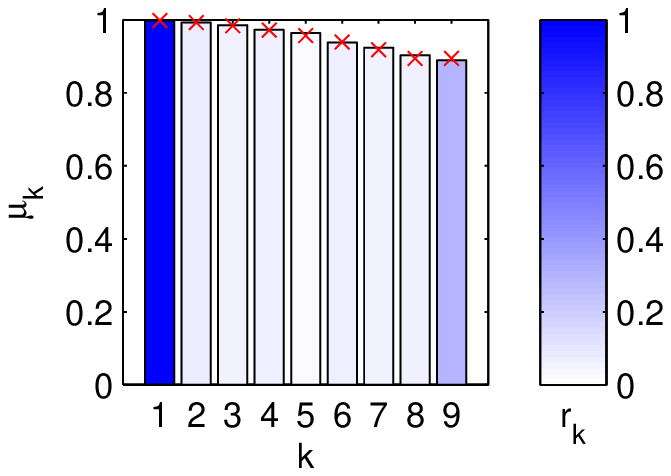}
\caption{}
\end{subfigure}
\caption{Data sets (top) and eigenvalue spectra from diffusion maps analysis (bottom) for strips with (a) $L_1 = 2$, $L_2 = 1$, (b) $L_1 = 4$, $L_2 = 1$, (c) $L_1 = 8$, $L_2 = 1$. The empirical eigenvalues are plotted in blue, and the analytical eigenvalues are plotted in red. The bars are colored by $r_k$, as defined in \eqref{eq:cv_error}. From the eigenvalues which are identified as parametrizing unique eigendirections (indicated by the darker blue bars), the estimated length ratio from \eqref{eq:est_lengths} is (a) 2.2, (b) 4.1, (c) 8.7.
}
\label{fig:strip_compare_analytic}
\end{figure}

We consider three different two-dimensional strip data sets.
Each data set contains $m=2000$ data points uniformly sampled from the strip.
Figure~\ref{fig:strip_compare_analytic} shows the data sets and the diffusion maps eigenspectra.
The eigenvalues are colored by the leave-one-out cross-validation error $r_k$ as defined in \eqref{eq:cv_error};
a small value of $r_k$ indicates that the corresponding eigenvector is a repeated eigendirection, while a
large value of $r_k$ indicates that the corresponding eigenvector describes a new direction in the data.

The eigenvalues are consistent with the known analytic eigenvalues of
the Laplacian (see \eqref{eq:evals} and \eqref{eq:evals_relationship}), shown in red.
Furthermore, the two unique eigendirections can easily be identified, since their corresponding regression error $r_k$ is large.
As expected, the gap between the two meaningful eigenvalues increases as the strip becomes longer.
Using \eqref{eq:est_lengths}, we can accurately estimate the relative lengths of the two unique eigendirections in each data set.

\subsubsection{Swiss roll}

\begin{figure}[!t]
\begin{subfigure}{0.2\textwidth}
\centering
\includegraphics[width=\textwidth]{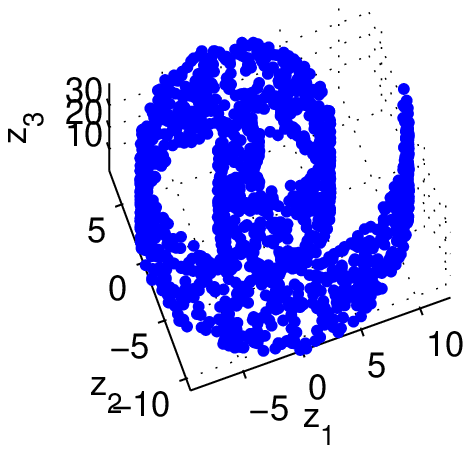}
\caption{}
\label{subfig:swissroll1}
\end{subfigure}
\begin{subfigure}{0.25\textwidth}
\centering
\includegraphics[width=\textwidth]{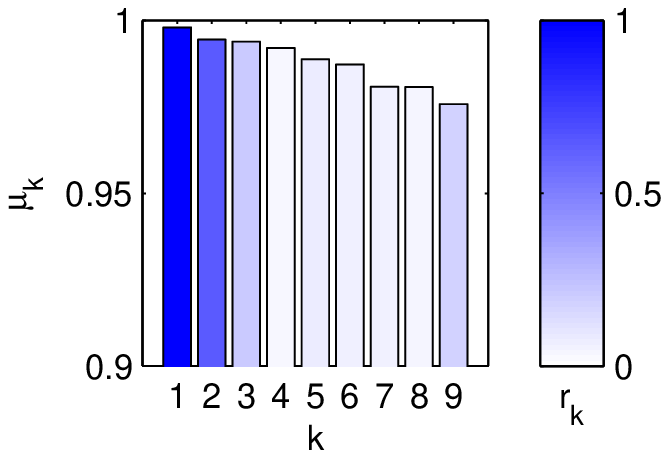}
\caption{}
\label{subfig:swissroll1_evals}
\end{subfigure}
\begin{subfigure}{0.25\textwidth}
\centering
\includegraphics[width=\textwidth]{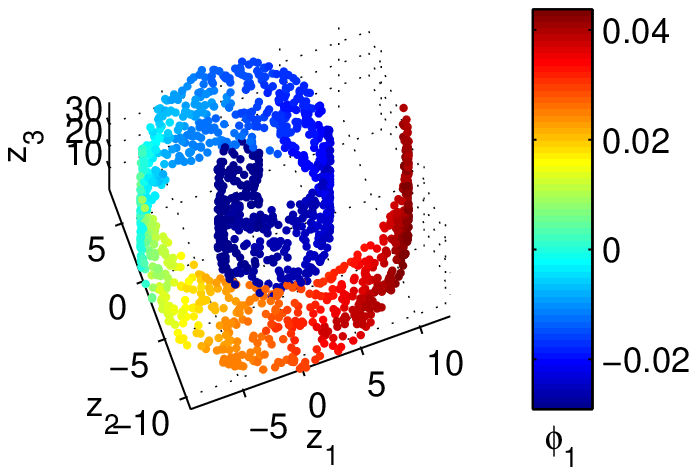}
\caption{}
\label{subfig:swissroll1_color1}
\end{subfigure}
\begin{subfigure}{0.25\textwidth}
\centering
\includegraphics[width=\textwidth]{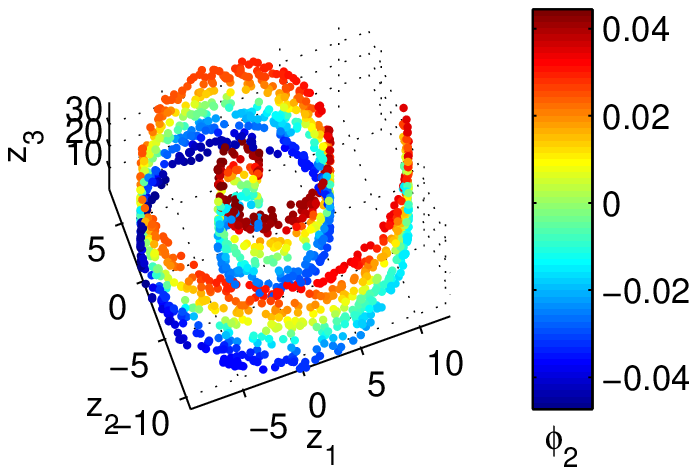}
\caption{}
\label{subfig:swissroll1_color2}
\end{subfigure}

\begin{subfigure}{0.2\textwidth}
\centering
\includegraphics[width=\textwidth]{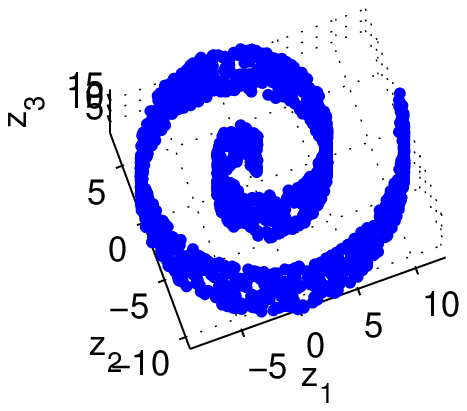}
\caption{}
\label{subfig:swissroll2}
\end{subfigure}%
\begin{subfigure}{0.25\textwidth}
\centering
\includegraphics[width=\textwidth]{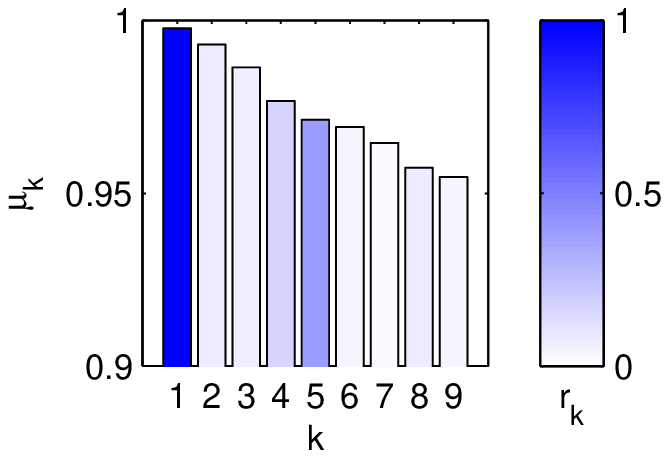}
\caption{}
\label{subfig:swissroll2_evals}
\end{subfigure}
\begin{subfigure}{0.25\textwidth}
\centering
\includegraphics[width=\textwidth]{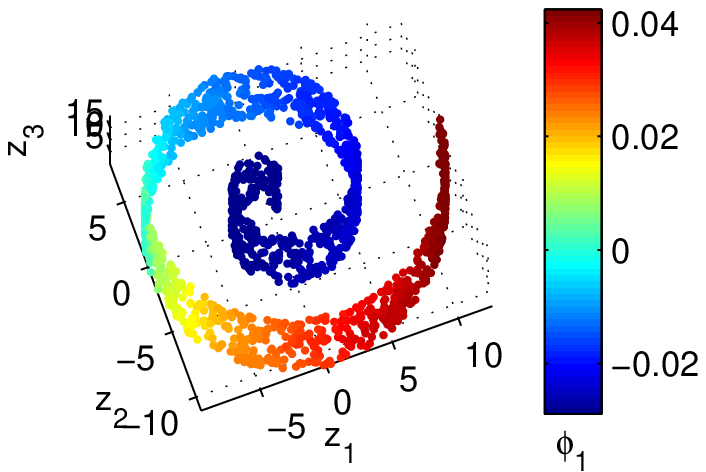}
\caption{}
\label{subfig:swissroll2_color1}
\end{subfigure}
\begin{subfigure}{0.25\textwidth}
\centering
\includegraphics[width=\textwidth]{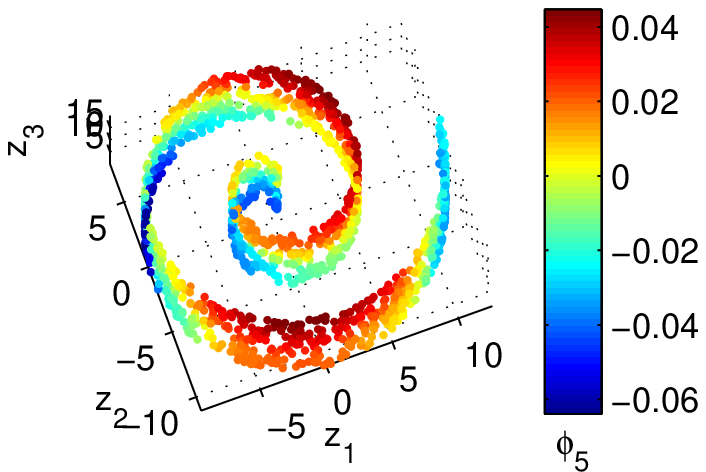}
\caption{}
\label{subfig:swissroll2_color2}
\end{subfigure}
\caption{Swiss roll example. (a) Data set 1; $h= 40$. (b) Eigenvalue spectrum from the diffusion maps analysis of data set 1. (c) Data set 1, colored by the first diffusion maps eigenvector. (d) Data set 1, colored by the second diffusion maps eigenvector. (e) Data set 2; $h = 20$. (f) Eigenvalue spectrum from the diffusion maps analysis of data set 2. (g) Data set 2, colored by the first diffusion maps eigenvector. (h) Data set 2, colored by the fifth diffusion maps eigenvector. }
\label{fig:swiss_rolls}	
\end{figure}

Our second illustrative example consists of two different Swiss roll data sets.
The data are sampled according to
\begin{equation}
(z_1, z_2, z_3) = (\theta \cos \theta , \theta \sin \theta, h t)
\end{equation}
where $\theta$ is sampled such that $z_1, z_2$ are uniformly sampled along the arclength of the spiral, and $t$ is uniformly sampled from $[0,1]$.
Note that, in this example, $z_1, z_2, z_3$ are the original data coordinates; however, the data lie on a
two-dimensional manifold parametrized by the $\theta$ and $t$.
The height of the first Swiss roll is $h = 40$, while the height of the second is $h = 20$.
Each data set consists of $m=1500$ points, shown in Figures~\ref{subfig:swissroll1}~and~\ref{subfig:swissroll2}.
%

Figures~\ref{subfig:swissroll1_evals}~and~\ref{subfig:swissroll2_evals} show the eigenvalue spectra from the analysis of the two data sets.
Similar to Figure~\ref{fig:strip_compare_analytic}, the bars are colored by the leave-one-out cross-validation error, 
where a small value of $r_k$ indicates that the corresponding eigenvector is a repeated eigendirection.
From these plots, one can conclude that the first two eigenvectors $\phi_1$ and $\phi_2$ parametrize the first data set, 
while $\phi_1$ and $\phi_5$ parametrize the second.
Figures~\ref{subfig:swissroll1_color1},\ref{subfig:swissroll1_color2},\ref{subfig:swissroll2_color1},~and~\ref{subfig:swissroll2_color2} 
show the two data sets, colored by the two eigenvectors identified as parametrizing the unique eigendirections.
As expected, these eigenvectors are one-to-one with the arclength along the spiral, and the height of the Swiss roll.

\subsubsection{Torus}

\begin{figure}[t]
\centering
\begin{subfigure}{1.5in}
\centering
\includegraphics[height=0.75in]{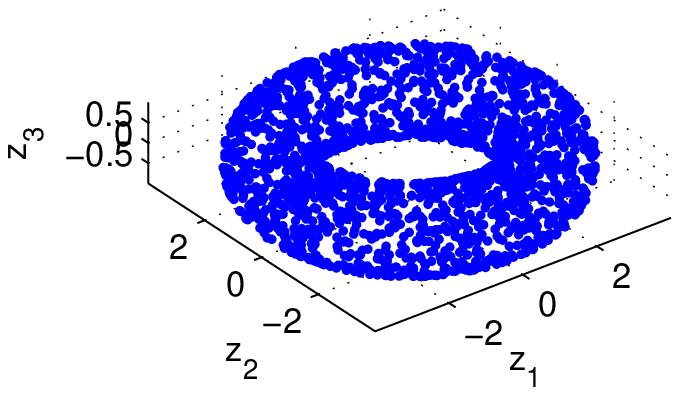}
\includegraphics[height=1in]{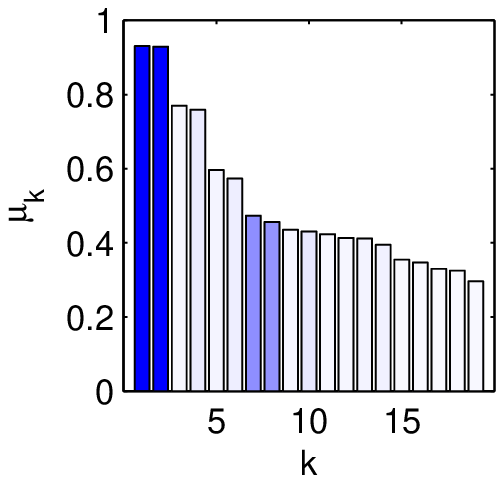}
\caption{}
\end{subfigure}
%
%
\begin{subfigure}{1.5in}
\centering
\includegraphics[height=0.75in]{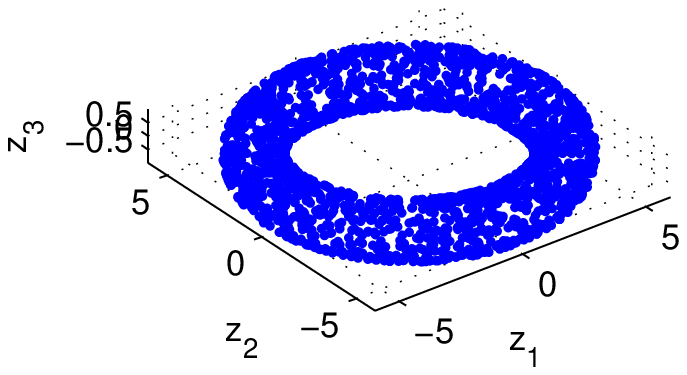}
\includegraphics[height=1in]{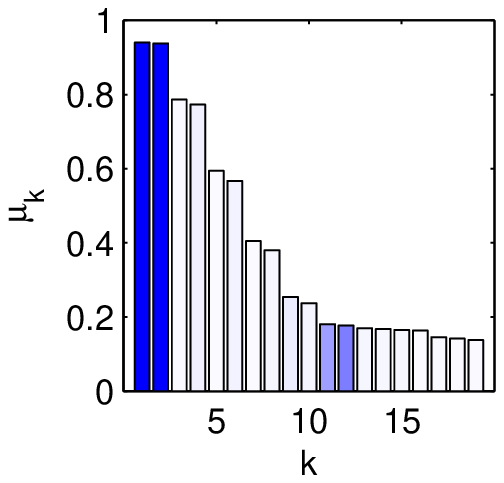}
\caption{}
\end{subfigure}
%
%
\begin{subfigure}{1.5in}
\centering
\includegraphics[height=0.75in]{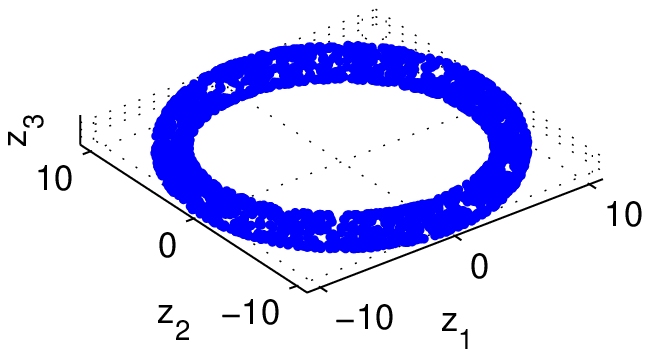}
\includegraphics[height=1in]{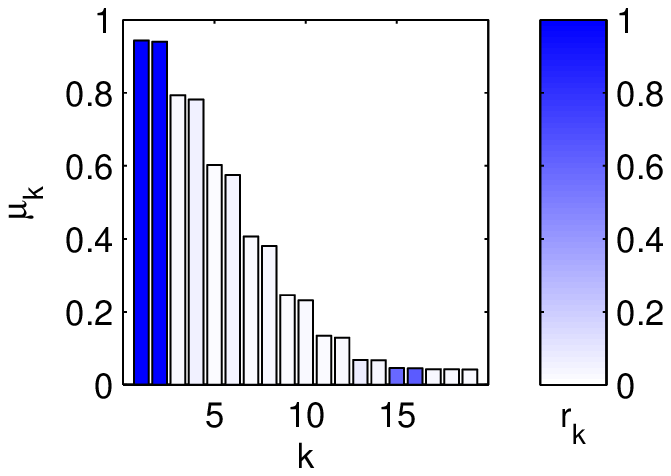}
\caption{}
\end{subfigure}
\hfill
\caption{Tori data sets (top, see \eqref{eq:torus}) and corresponding diffusion maps eigenvalues (bottom) for (a) $r_1 = 3$, $r_2 = 1$. (b) $r_1 = 5$, $r_2 = 1$. (c) $r_1 = 10$, $r_2 = 1$. The tori are parametrized by two angles: $\theta_1$, which parametrizes the large outer circle, and $\theta_2$, which parametrizes the smaller inner circle. In all three data sets, the first two eigenvalues/eigenvectors correspond to $\sin \theta_1$ and $\cos \theta_1$. The second pair of unique eigendirections (corresponding to $\sin \theta_2$ and $\cos \theta_2$) are captured by components 7 and 8, 11 and 12, and 15 and 16, respectively.}
\label{fig:torus}
\end{figure}

For the third example, we consider a torus defined by
\begin{equation} \label{eq:torus}
\left( z_1, z_2, z_3 \right) = \left(  
(r_1 + r_2 \cos \theta_2 ) \cos \theta_1, 
(r_1 + r_2 \cos \theta_2 ) \sin \theta_1, 
r_2 \sin \theta_2
\right) 
\end{equation}
where $r_1 > r_2$ are the outer and inner radii, respectively, and $\theta_1$ and $\theta_2$ are uniformly and independently sampled from $[0, 2\pi)$.
In this example, $z_1, z_2, z_3$ are the coordinates of the data; however, we expect to obtain from our analysis 
two eigenfunctions ($\sin \theta_1$ and $\cos \theta_1$) which parametrize the outer circle, and 
two eigenfunctions ($\sin \theta_2$ and $\cos \theta_2$) which parametrize the inner circle.
Intuitively, increasing $r_1$ can be viewed as analogous to increasing the ratio of $L_1$ to $L_2$ in the 
strip example and makes the eigendirections which parametrize the outer circle ($\cos \theta_1$ and $\sin \theta_1$) more 
dominant compared to the eigendirections with parametrize the inner circle.
Figure~\ref{fig:torus} shows the eigenspectra from the analysis of three different tori, colored again by the leave-one-out cross-validation error $r_k$.
We observe that as the torus becomes thinner ($r_1$ increases), the second pair of eigenvalues corresponding to unique eigendirections (corresponding to $\sin \theta_2$ and $\cos \theta_2$) moves farther down in the spectrum.

\section{Chemotaxis: A Case Study}

Our main motivation for identifying the unique eigendirections is to facilitate the analysis of 
complex data sets where the underlying dimensionality and structure is not readily apparent.
In data from complex dynamical systems, noisy microscopic behavior often gives rise to coherent macroscopic dynamics.
In general, this mapping from microscale to macroscale is not always obvious.
We will show how we can use our proposed methodology to extract a parametrization of data collected 
at the microscale which is consistent with a certain macroscopic behavior, without any {\em a priori} knowledge of the appropriate 
microscopic or macroscopic model.
Furthermore, determining the {\em true} dimensionality of such a parametrization by identifying 
the unique eigendirections reveals the requisite dimensionality of a reduced model which captures the relevant macroscopic behavior.
Knowing appropriate macroscopic variables and the true dimensionality can help inform modeling efforts and aid in 
positing useful macroscale models.

Our model problem describes the process of cellular chemotaxis \cite{othmer2000diffusion}, 
where biological cells exhibit coherent macroscopic dynamics regulated by extracellular sensed signals 
in order to accomplish tasks such as finding food or navigating away from toxins.
Several microscopic models have been proposed to describe chemotactic dynamics \cite{othmer1988models, codling2008random}.
We will analyze one such model described by a one-dimensional velocity jump process \cite{othmer2000diffusion}.
This specific example has an analytic macroscopic description in which the 
dynamics of a large ensemble of cells depend on the value of a single system parameter.
This description serves as a ``ground truth" and will allow us to validate our results which are obtained in an unsupervised manner.

Thus far, we only considered synthetic data sets for which the Euclidean distance between data points served as an informative metric.
Here, we will show that our approach, {\em when utilizing the appropriate statistical observers and affinity metric between pairs of observations}, 
uncovers a parametrization of the microscopic data which is consistent with the macroscopic model.
Furthermore, we will show that changes in the ``mode" of dynamical behavior resulting from changes in the system parameters can be automatically detected.

\subsection{Problem description}

The microscopic model consists of a collection of $N$ cells (for our simulations, we take $N=1000$) whose states are 
defined by their positions and velocities on a line, and the dynamics of each cell are governed by a stochastic process.
Let $x_i(t)$ and $v_i(t)$ denote the position and velocity, respectively, of cell $i$ at time $t$.
The velocity of each cell is $\pm s$, where $s$ is a (fixed) speed.
We initialize the cells such that
\begin{equation}\label{eqn:system}
\begin{aligned}
x_i(0) & = 0 \\
\mathbb{P} \{ v_i(0) = +s \} & = p
\end{aligned}
\end{equation}
where $0 < p < 1$ is the probability of a cell initially moving to the right.
The velocity of each cell randomly switches between $\pm s$ following an (independent) Poisson process with rate $\lambda$ (this switching is physically controlled by extracellular signals).
For our specific simulations, we set $s^2/\lambda=1$, which is consistent with the analysis presented in \cite{othmer1988models}.
We note that we have chosen a very specific one-parameter family of initial conditions which lead to simple dynamics and allow us to illustrate our main points;
however, our analysis is not restricted to such specific cases and more complex sets of initial conditions could be used.
Each data set consists of $10$ stochastic simulations, with initial conditions uniformly chosen such that $0.1 \le p  \le 0.9$.
We allow each simulation to evolve for $t_{max}$ time units, and record the states of the $N$ cells every $\Delta t$ time units; we use data with $t > 0$ for analysis.
We will show that the parameter $t_{obs} \equiv t_{max}/N$, which determines the time scale of observation, is of critical importance when we address the limiting asymptotic behavior.

\begin{figure}[t!]
\def \figwidth {0.25\textwidth}
\centering
\begin{subfigure}{\figwidth}
\includegraphics[width=\textwidth]{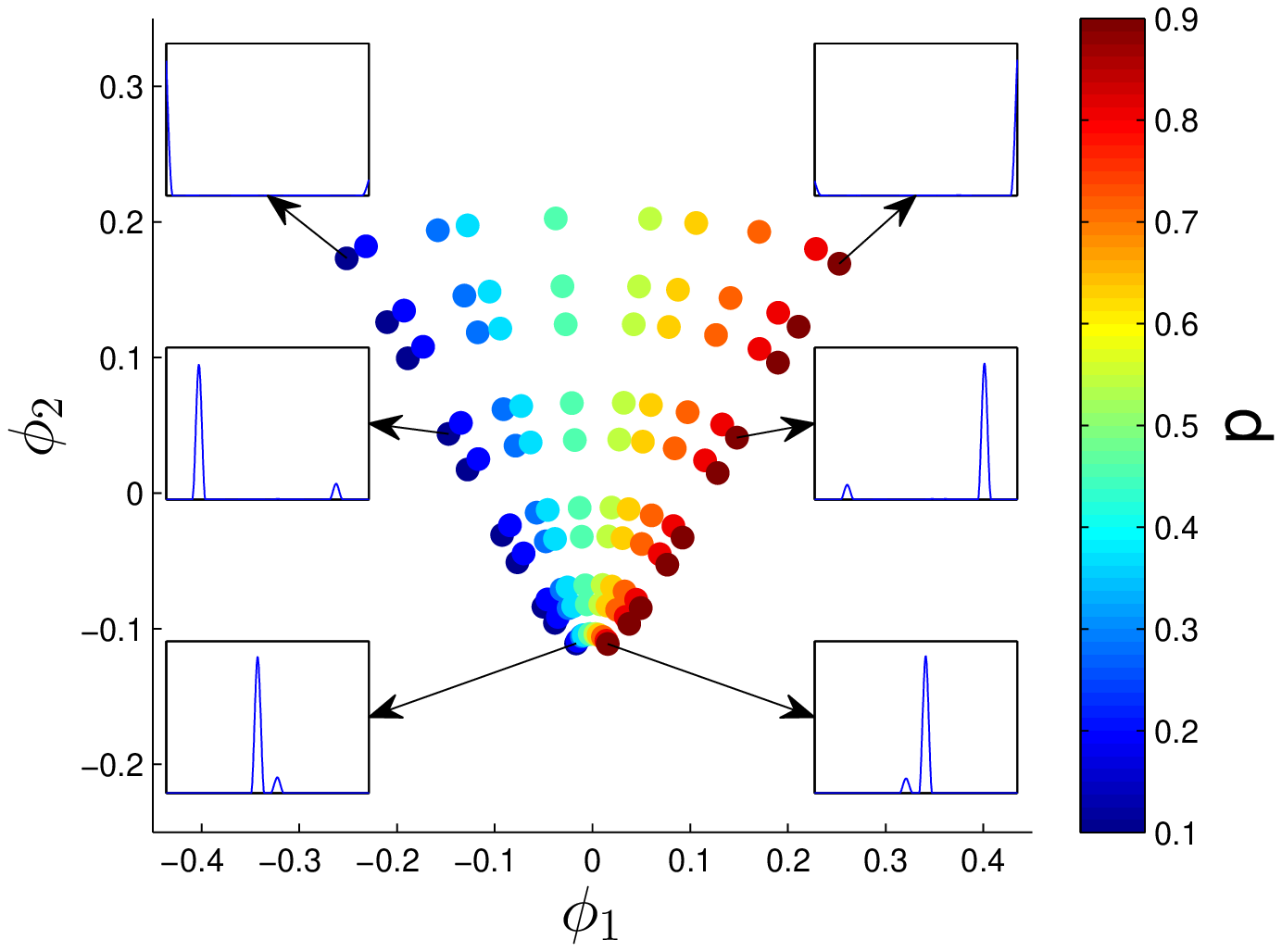}
\caption{}
\label{subfig:small_lambda_p}
\end{subfigure}
\begin{subfigure}{\figwidth}
\includegraphics[width=\textwidth]{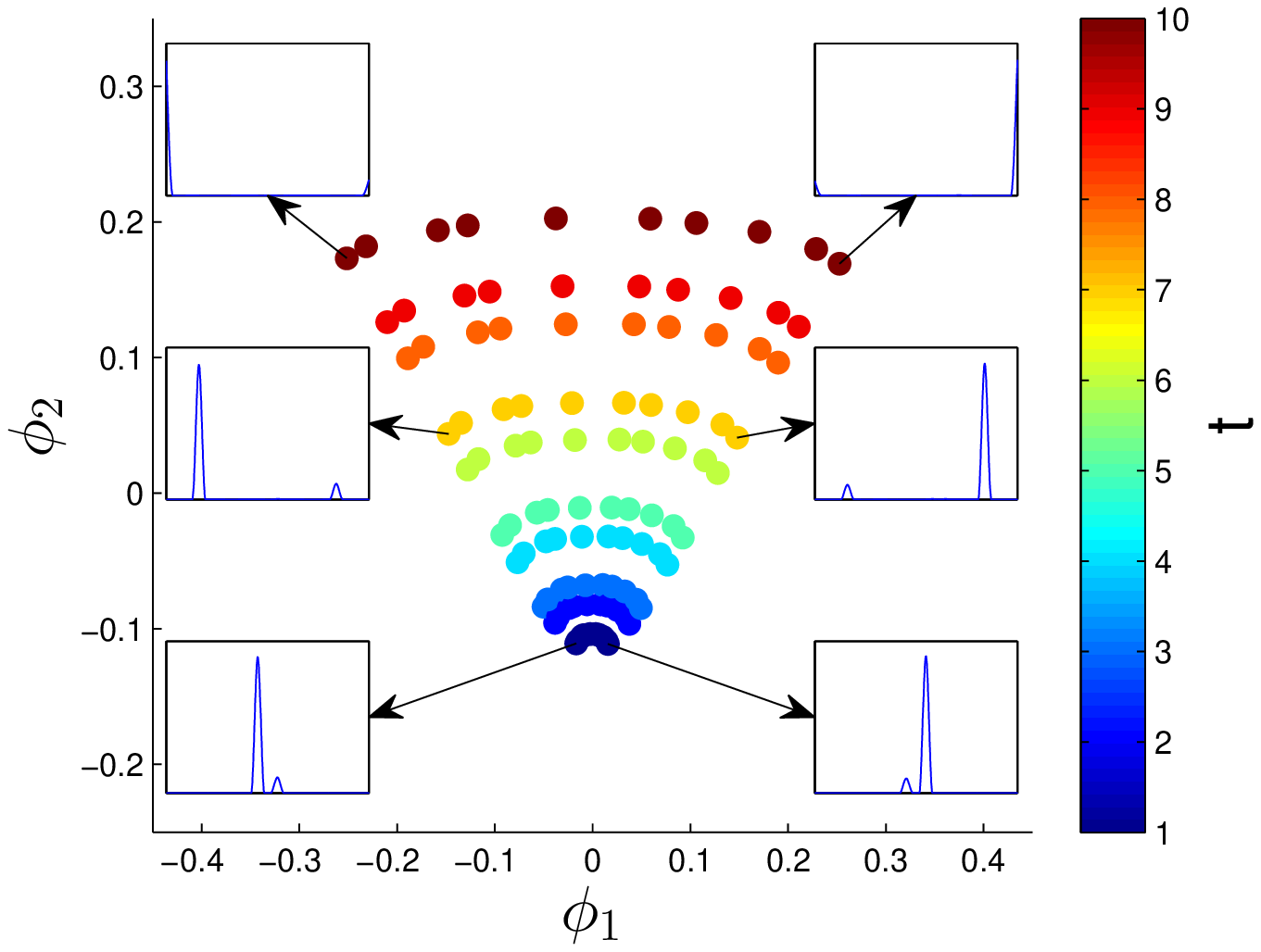}
\caption{}
\label{subfig:small_lambda_t}
\end{subfigure}
\begin{subfigure}{\figwidth}
\includegraphics[width=\textwidth]{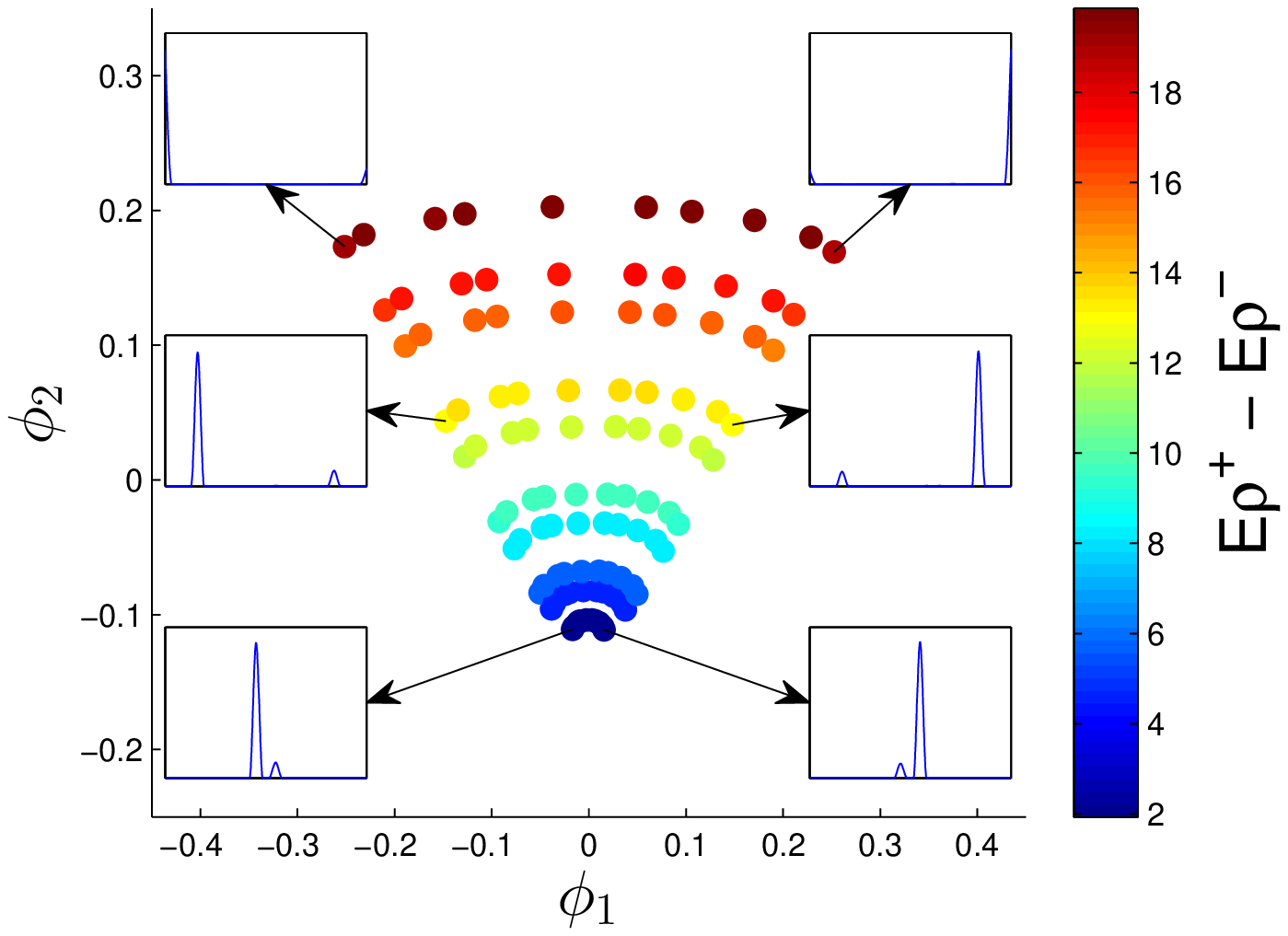}
\caption{}
\label{subfig:small_lambda_rho}
\end{subfigure}

\begin{subfigure}{\figwidth}
\includegraphics[width=\textwidth]{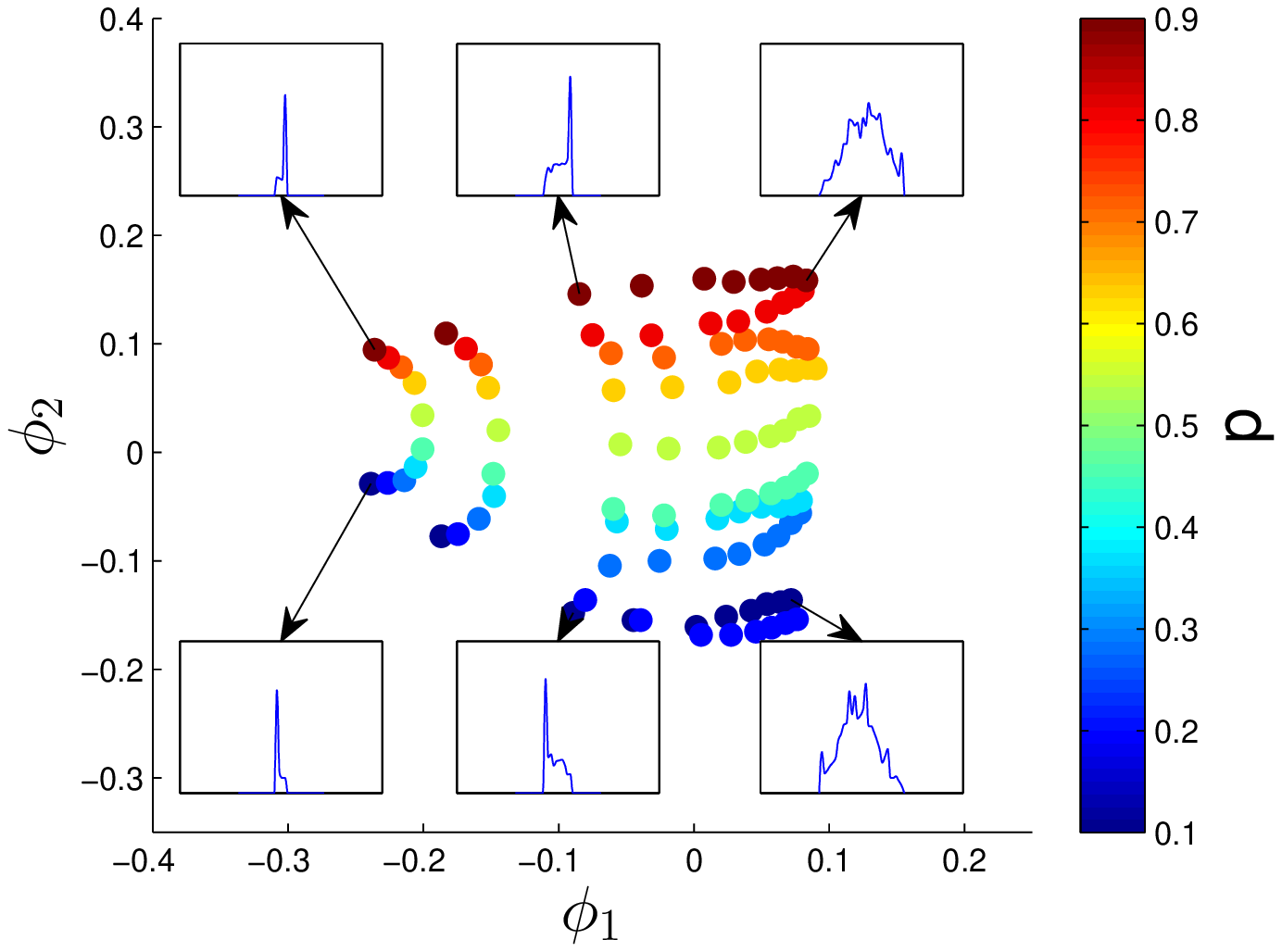}
\caption{}
\label{subfig:large_lambda_p}
\end{subfigure}
\begin{subfigure}{\figwidth}
\includegraphics[width=\textwidth]{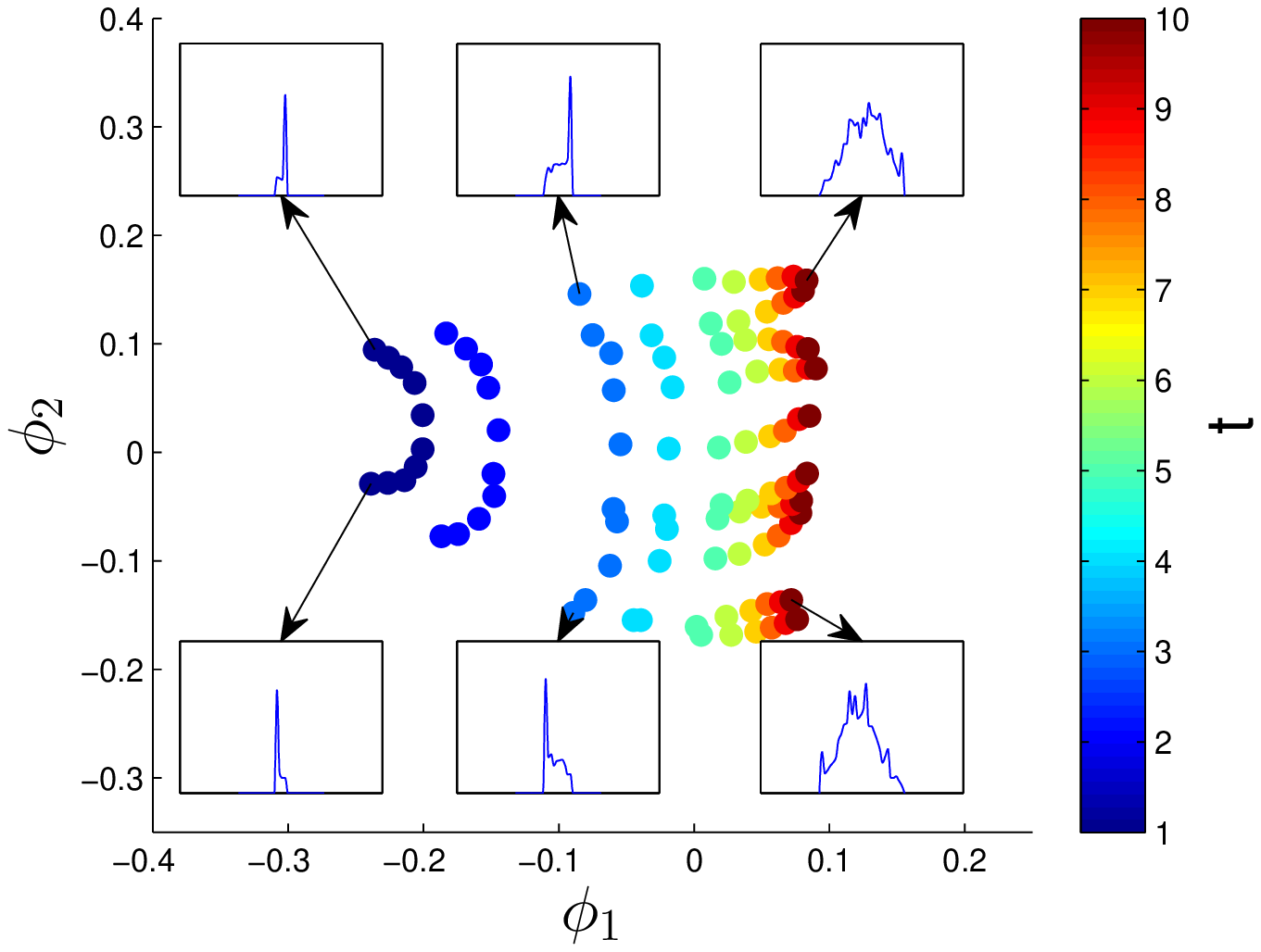}
\caption{}
\label{subfig:large_lambda_t}
\end{subfigure}
\begin{subfigure}{\figwidth}
\includegraphics[width=\textwidth]{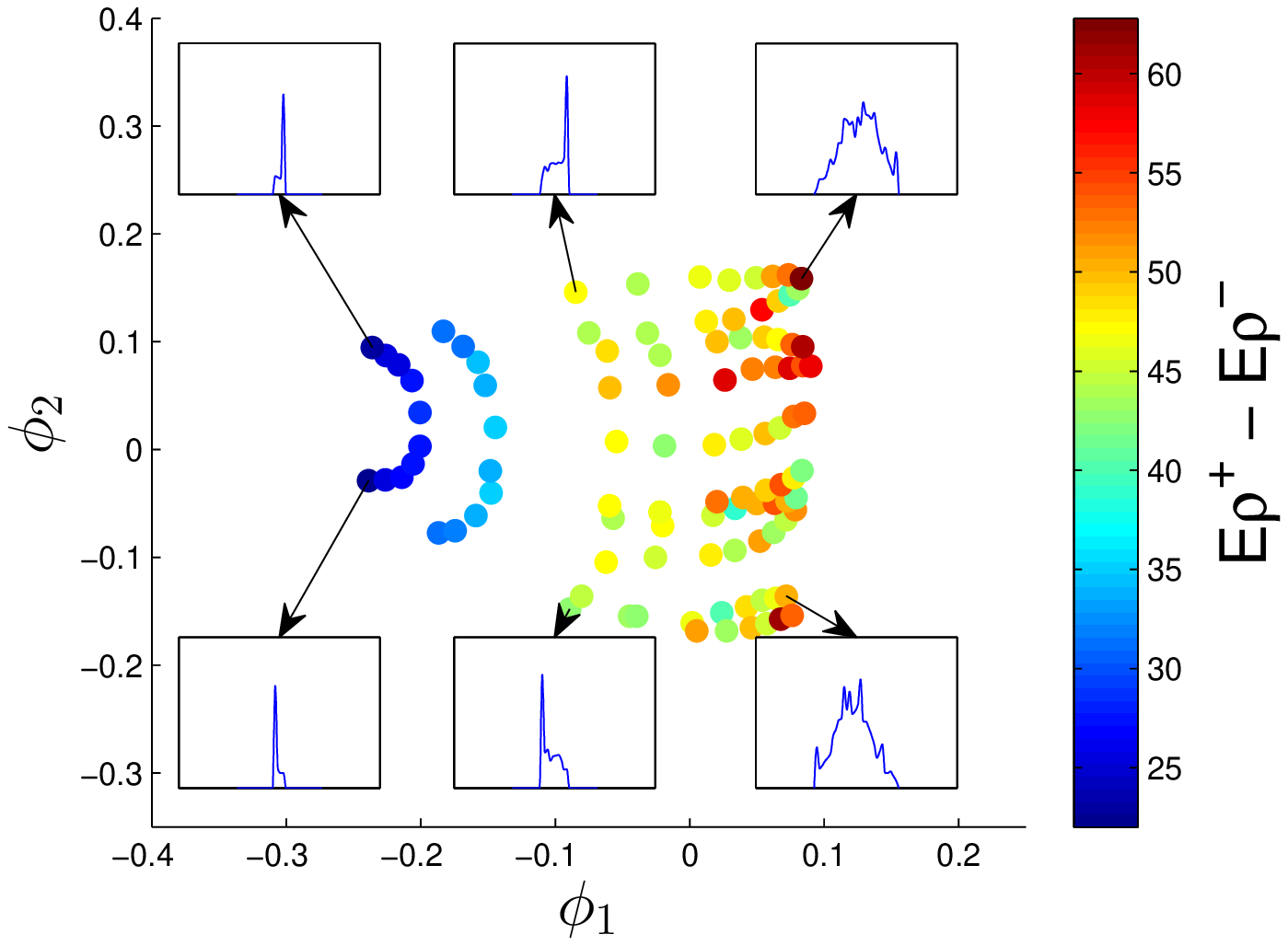}
\caption{}
\label{subfig:large_lambda_rho}
\end{subfigure}
\caption{Embeddings using the first two diffusion maps eigenvectors computed from simulation data of the velocity jump process with $t_{max} = 10$, $\Delta t=1$, and (a--c) $\lambda=1$, $s=1$, (d--f) $\lambda=400$, $s=20$.  The distances used in the diffusion maps kernel are the earth mover's distances between the histograms of cell positions. The data are colored by (a, d) $p$, the initial probability of a cell moving to the right, (b, e) $t$, time, and (c, f) $E \rho^+ - E \rho^-$, the difference between the average position of the left- and right-moving cells. Representative histograms of the cell positions are shown for selected data points. }
\label{fig:dmaps_embed_emd}
\end{figure}

\subsection{Observers and metrics}

In general, manifold learning techniques have two essential components.
One is the appropriate {\em observers} of the system state.
These observers should be informative as to the state of the system, as well as insensitive to noise in the system.
The second is a {\em distance metric} between the observations that captures a notion of locality: 
observations which we perceive to be similar should yield a small value for this distance.

For the chemotaxis example, we use histograms of the cell positions as observers.
Histograms are invariant to the indexing of the cells, while retaining information about the relative spatial locations of the cells.
Histograms are also robust to noise \cite{talmon2013empirical}.
Instead of the standard Euclidean distance, we use the earth mover's distance (EMD) \cite{rubner2000earth} as the metric between pairs of histograms.
Conceptually, the EMD measures how much ``work'' it takes to transform one probability density into another.
It therefore not only considers where the densities are inconsistent, but also how far apart the inconsistencies are from each other.
Although the brute-force computation of the EMD is computationally expensive, 
there has been a plethora of work in developing efficient algorithms for its approximation \cite{Pele-eccv2008, Pele-iccv2009, leeb2014lipschitz}.
For the specific case of spatially one-dimensional data, the EMD is equivalent to the $L_1$-norm between the 
cumulative distribution functions of the data \cite{rubner2000perceptual}, which can be approximated from histograms as
\begin{equation}
\| z(i) - z(j) \|_{EMD} \approx \sum_{l=1}^{n} \left| \sum_{k=1}^l z_k(i) - \sum_{k=1}^l z_k(j) \right|;
\end{equation}
the histograms $z(i), z(j)$ are defined on $n$ equally-spaced bins in $\mathbb{R}$ (for our simulations, we will take $n=32$), and $z_k$ denotes the $k^{th}$ bin.
Here, we choose to only use the positions of the cells in the distance computation; utilizing both the positions and velocities produces qualitatively similar results (not shown).
We note that although histograms are not essential to the theory of EMD, they make the computation practical.

\subsection{Results}

\subsubsection{Identifying the unique eigendirections}

\begin{figure}[t]
\centering
\def\figheight{1.3in}
\begin{subfigure}[t]{1.5in}
\centering
\includegraphics[height=\figheight]{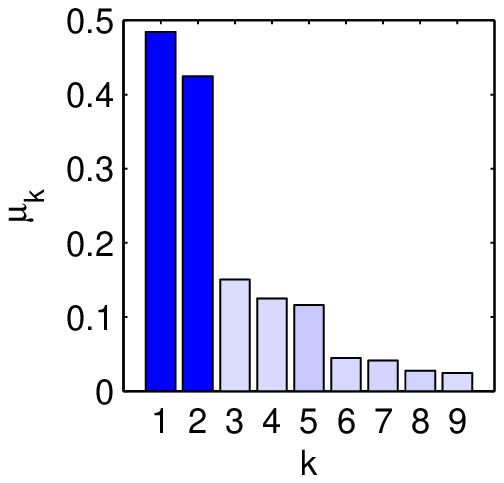}
\includegraphics[height=\figheight]{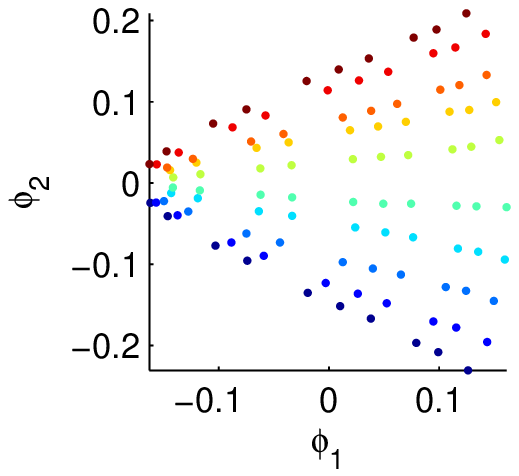}
\vspace{1.2in}
\caption{}
\end{subfigure}
\begin{subfigure}[t]{1.5in}
\centering
\includegraphics[height=\figheight]{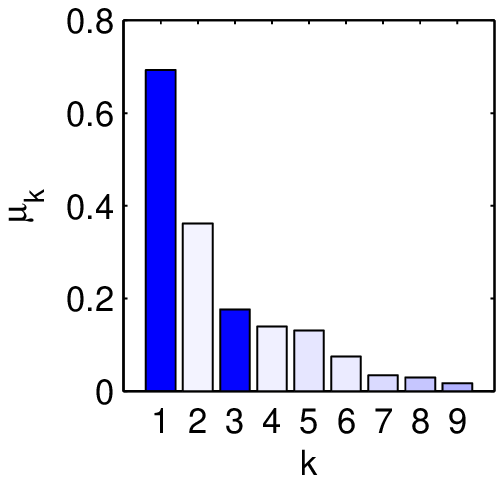}
\includegraphics[height=\figheight]{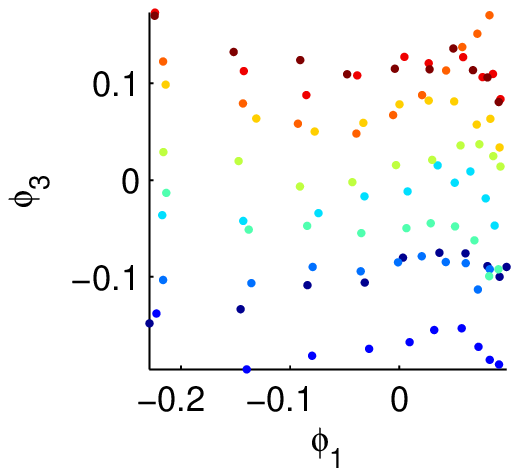}
\includegraphics[height=\figheight]{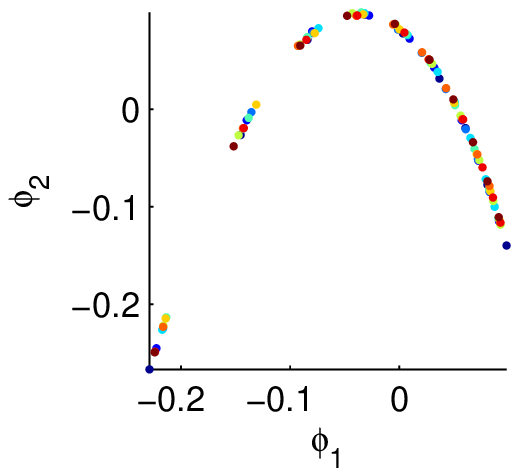}
\caption{}
\end{subfigure}
\begin{subfigure}[t]{2in}
\centering
\includegraphics[height=\figheight]{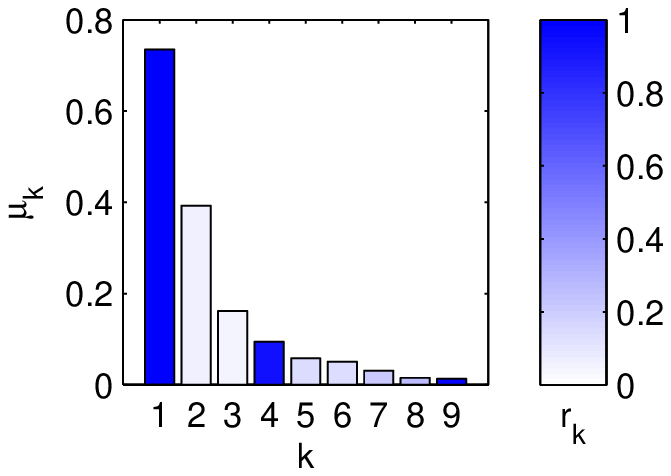}
\includegraphics[height=\figheight]{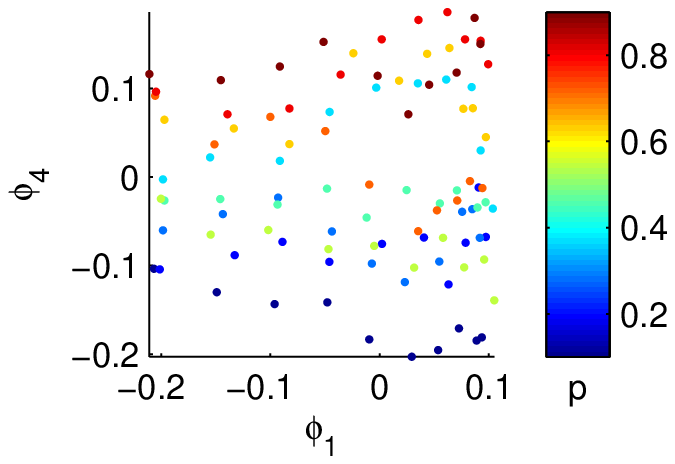}
\includegraphics[height=\figheight]{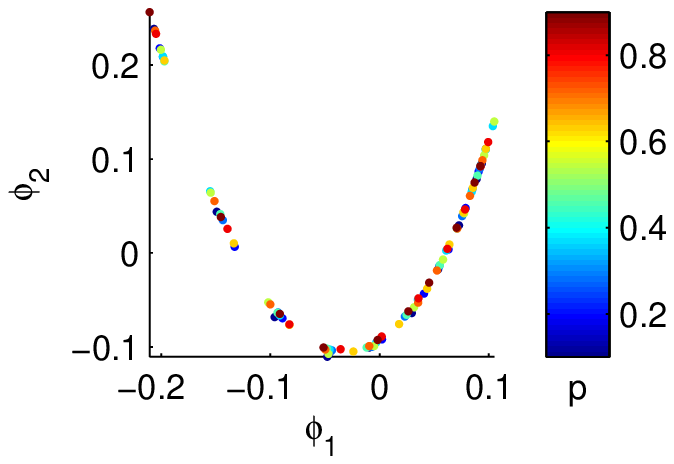}
\caption{}
\end{subfigure}
\caption{Analysis of three different chemotaxis simulation data sets. (a) $\lambda = 100$, $s = 10$. 
(b) $\lambda = 1600$, $s = 40$. (c) $\lambda = 6400$, $s = 80$. We set $t_{max} = 10$ and $\Delta t=1$, and the 
distances used in the diffusion maps kernel are the earth mover's distances between the histograms of cell positions. 
For each data set, the eigenvalue spectrum, colored by $r_k$, is shown in the top row. 
From the spectra, we can see that the first two components are informative for (a), that components 1 and 3 are informative for (b), 
and that components 1 and 4 are informative for (c). 
The corresponding reduced diffusion maps embeddings are shown in the middle row. For comparison, 
the standard diffusion maps embeddings using the first two components are also shown in the bottom row in (b) and (c).}
\label{fig:chemotaxis_simulations_harmonics}
\end{figure}

Figure~\ref{fig:dmaps_embed_emd} shows the results of analyzing two sets of chemotaxis simulations.
One set of simulations (Figures~\ref{subfig:small_lambda_p}--\ref{subfig:small_lambda_rho}) is for a relatively small value of $\lambda$, 
while the other set (Figures~\ref{subfig:large_lambda_p}--\ref{subfig:large_lambda_rho}) is for a larger value of $\lambda$.
For both sets of simulations, the macroscopic variables $p$, which controls the initial distribution of cell velocities, and $t$, the time, 
are well-correlated with the eigenvectors $\phi_1$ and $\phi_2$.
In particular, $p$ characterizes the fraction of initially right-moving cells (alternatively, the initial {\em flux} of the right-moving cells). 
%
%
The dominant coordinate $\phi_1$ is correlated with $p$ for the small $\lambda$ case (Figure~\ref{subfig:small_lambda_p}), 
and correlated with $t$ for the large $\lambda$ case (Figure~\ref{subfig:large_lambda_t}), indicating that the relative 
importance of $p$ and $t$ changes in the two simulations.
%

As illustrated in the synthetic data sets we discussed previously, the two leading eigenvectors are not 
guaranteed to correspond to unique eigendirections.
Figure~\ref{fig:chemotaxis_simulations_harmonics} shows the results of analyzing three different simulations 
which span a larger range of $\lambda$ values compared to Figure~\ref{fig:dmaps_embed_emd}.
We see from the eigenspectra and the cross-validation errors $r_k$ that the second unique eigendirection moves 
down in the spectrum as $\lambda$ increases.
The two-dimensional reduced diffusion maps embeddings obtained from eigenvectors which parametrize unique eigendirections 
capture the macroscopic variables $p$ and $t$ (we only show the correspondence of the two identified unique eigendirections 
with $p$ in the middle row of Figure~\ref{fig:chemotaxis_simulations_harmonics}; the correspondence 
between these eigendirections and $t$ is similarly consistent).
In contrast, using the two leading eigenvectors $\phi_1$ and $\phi_2$ produces 
uninformative embeddings for large values of $\lambda$, as $\phi_2$ is a repeated eigendirection which 
does not parametrize any new directions in the data.
Furthermore, although the eigenvalue spectra do not exhibit any large spectral gaps, 
the gap between the eigenvalues which correspond to new directions in the data (indicated by the dark blue) increases with $\lambda$, 
indicating that the underlying manifold becomes narrower, and the corresponding eigenvectors 
provide us with an informative two-dimensional parametrization of the data. 
It now becomes clear that looking at the eigenvectors {\em modulo} repeated eigendirections is essential for extracting an informative parametrization of the data, as well as characterizing the effective dimensionality of the data.

\subsubsection{Analytic macroscopic description}

This particular model system can be usefully described through a known analytic macroscopic equation that governs the overall system behavior.
For a large collection of cells ($N \rightarrow \infty$), the system can be described using the evolution of the cell density.
Let $\rho(x, t)$ denote this evolving cell density field, and let $\rho^-(x, t)$ and $\rho^+(x, t)$ denote the 
densities of the left-moving and of the right-moving cells, respectively.
It can be shown that, as $N \rightarrow \infty$, the densities satisfy the following set of coupled partial differential equations (PDEs) \cite{othmer2000diffusion}:
\begin{equation} \label{eqn:coupled_pdes}
\begin{aligned}
\frac{\partial \rho^+}{\partial t} + s \frac{\partial \rho^+}{\partial x} & = -\lambda \rho^+ +\lambda \rho^- \\
\frac{\partial \rho^-}{\partial t} - s \frac{\partial \rho^-}{\partial x} & = \lambda \rho^+ -\lambda \rho^- .
\end{aligned}
\end{equation}
Alternatively, \eqref{eqn:coupled_pdes} can be rewritten as a single, second--order PDE (the telegrapher equation):
\begin{equation} \label{eq:second_order_pde}
\frac{\partial^2 \rho}{\partial t^2} + 2 \lambda \frac{\partial \rho}{\partial t} = s^2 \frac{\partial ^2 \rho}{\partial x^2} .
\end{equation}
From \eqref{eq:second_order_pde}, we see that, for fixed values of $\lambda$ and $s$, the macroscopic 
state of the system (the probability density of the cells) is indeed a function of the initial condition (parametrized by $p$) and the time $t$.
This implies that for fixed $\lambda$ and $s$, the {\em microscopic} data in a high-dimensional ambient space (e.g., the positions of all $N$ cells) should lie on a two-dimensional manifold parametrized by $p$ (which controls the initial ratio of $\rho^+$ and $\rho^-$) and $t$ (which describes the evolution);
uncovering the low-dimensional structure of the microscopic data helps reveal this manifold, leading to a subsequent physically meaningful
parametrization of it.
This is consistent with the results presented in Figures~\ref{fig:dmaps_embed_emd}~and~\ref{fig:chemotaxis_simulations_harmonics}, where the two-dimensional embeddings obtained from microscopic simulation data are one-to-one with $p$ and $t$.

Guided by the known theory of the model macroscopic PDE system, we consider two asymptotic regimes for the simulation.
When $\lambda \rightarrow 0$, the right-hand side of \eqref{eqn:coupled_pdes} tends to 0, and \eqref{eqn:coupled_pdes} becomes two uncoupled wave equations,
\begin{equation}
\begin{aligned}
\frac{\partial \rho^+}{\partial t} + s \frac{\partial \rho^+}{\partial x} & = 0 \\
\frac{\partial \rho^-}{\partial t} - s \frac{\partial \rho^-}{\partial x} & = 0.
\end{aligned}
\end{equation}
When $\lambda \rightarrow \infty$ with $s^2/\lambda$ fixed, \eqref{eq:second_order_pde} approaches the diffusion equation,
\begin{equation} \label{eq:diff_eqn}
2 \frac{\partial \rho}{\partial t} = D \frac{\partial ^2 \rho}{\partial x^2},
\end{equation}
where $D=s^2/\lambda$.
The above analysis shows that the initial distribution of velocities of the cells 
(determined by $p$ in the microscopic simulations) plays a very different role depending on the value of $\lambda$.
When $\lambda \rightarrow 0$, the dynamics are described by the telegrapher equation, and the effect of the initial velocity distribution 
persists throughout the trajectory.
When $\lambda \rightarrow \infty$, the dynamics are described by a single diffusion equation, and the initial conditions 
for the velocity are not especially important, since the velocity distribution quickly equilibrates and we see purely diffusive behavior.

The relative importance of $p$ and $t$ from the analytic description are consistent with the results 
in Figure~\ref{fig:dmaps_embed_emd}, where $p$ is correlated with the dominant diffusion map coordinate 
when $\lambda$ is small, and becomes correlated with the subdominant coordinate when $\lambda$ is large.
Furthermore, in the small $\lambda$ regime (wave equation), shown in 
Figures \ref{subfig:small_lambda_p}--\ref{subfig:small_lambda_rho}, the points corresponding to 
small times are more tightly clustered than the points corresponding to large times.
This is in agreement with the macroscopic model: at small times, the cells are more 
condensed around $x=0$, and it is more difficult to distinguish the cells moving to the left from the cells moving to the right.
%
On the other hand, at large times, once the cells move away from the origin, this separation between left-moving and
right-moving cells is clear.
For the large $\lambda$ case (diffusion equation), shown in Figures \ref{subfig:large_lambda_p}--\ref{subfig:large_lambda_rho}, 
we observe that at small times (before the cell velocities have sufficient opportunity to switch and equilibrate), the initial velocity distribution $p$ can be well perceived in the
embedding in Figure \ref{subfig:large_lambda_p}, as the initial imbalance in left-right velocities affects
the initial displacements.
On the other hand, for longer times, we observe that the initial distribution $p$ cannot be well perceived
in the embedding in Figure \ref{subfig:large_lambda_p}, 
as the velocities have equilibrated and their initial distribution is practically forgotten.
Overall, in both cases, we obtain, in an unsupervised data-driven manner, a useful low-dimensional embedding of the data.
Correlating this embedding with potential candidate macroscopic observables (such as time, densities, fluxes, and initial condition statistics) helps elucidate
an accurate picture of the macroscopic variables that govern the system dynamics.

The analytic macroscopic model suggests that when $\lambda$ is small, the system dynamics 
can be described by the two densities $\rho^+$ and $\rho^-$, 
and when $\lambda$ is large, the dynamics can be described by a single density $\rho = \rho^+ + \rho^-$.
Figures~\ref{subfig:small_lambda_rho}~and~\ref{subfig:large_lambda_rho} show the data, 
colored by the difference in the average position of the left- and right-moving cells.
Clearly, for small $\lambda$, both $\rho^+$ and $\rho^-$ are required to describe the dynamics, and
the difference between the two densities is a useful observable in parametrizing the data manifold
(and thus a good macroscopic descriptor, a good ``candidate macroscopic variable").
However, for large $\lambda$, the two densities rapidly equilibrate, and the difference between the two 
densities is {\em not} a useful observable in parametrizing the data manifold. 


This analytic macroscopic description allows us to emphasize the importance of using an appropriate distance metric.
We also analyzed the two sets of simulation data from Figure~\ref{fig:dmaps_embed_emd} using the standard 
Euclidean distance between the histograms to compute the distances in \eqref{eq:W}.
We empirically found that there is no appreciable correlation between the embedding 
coordinates and what we know to be meaningful macroscopic variables, $p$ and $t$ (the correlations between the embedding 
coordinates and the governing macroscopic variables are all less than 60\%, with the correlations for 
the small $\lambda$ case being less than 20\%) with this similarity measure.
In contrast, the correlations between the embedding coordinates and the macroscopic variables when using EMD in the diffusion maps 
calculation are all greater than 80\%.
Clearly, using a metric which meaningfully describes the distances between observations 
is essential for obtaining an informative parametrization of data.
We note again that the metrics we used in this work only employed observations of cell positions (and not of cell instantaneous
velocities); if the instantaneous velocities are readily measurable, they could be incorporated in a useful
and possibly even more informative metric for data analysis.

\subsubsection{Detecting changes in dimensionality}

\begin{figure}[t]
\centering
\includegraphics[width=0.5\textwidth]{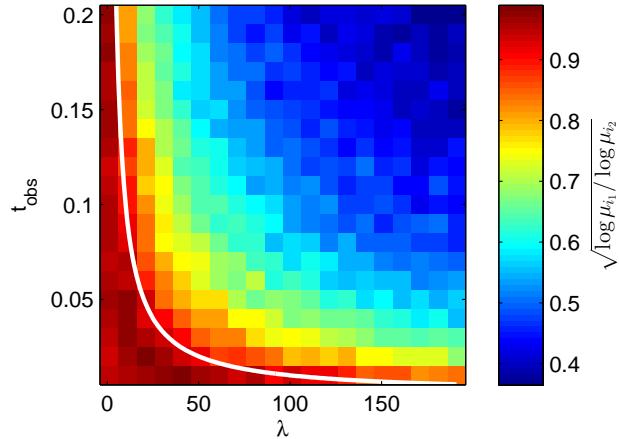}
\caption{Detecting the change in dimensionality. The appropriate ratio the two eigenvalues (see \eqref{eq:chemotaxis_eval_ratio}) which correspond to unique eigendirections is plotted as a function of $t_{obs}$ and $\lambda$. Each data point is the average eigenvalue ratio over three replicate data sets, and in each simulation, $\Delta t=t_{max}/10$. The distances used in the diffusion maps kernel are the earth mover's distances between the histograms of cell positions. The curve $t_{obs} = 1/\lambda$ is shown in white. }
\label{fig:chemotaxis_compare_timescales_evals}
\end{figure}

Detecting changes in effective dimensionality is essential for the analysis and modeling of (simulated or experimentally observed) dynamical systems.
The approximate dimensionally of a data set can be estimated by examining the 
two eigenvalues $\mu_{i_1} \ge \mu_{i_2}$ corresponding to unique eigendirections.
According to \eqref{eq:est_lengths}, we plot
\begin{equation}\label{eq:chemotaxis_eval_ratio}
 \sqrt{\frac{\log \mu_{i_1}}{\log \mu_{i_2}}} ;
\end{equation}
when this ratio becomes small, the data are {\em effectively} one-dimensional, as the extent of data along the second direction is very 
small compared with the first.

Figure~\ref{fig:chemotaxis_compare_timescales_evals} shows the eigenvalue ratio in \eqref{eq:chemotaxis_eval_ratio} as 
a function of $\lambda$ and the time scale of observation $t_{obs} = t_{max} / N$.
From this eigenvalue ratio, we can easily detect changes in the estimated dimensionality as we vary the relevant parameters.
For small $\lambda$ and/or short $t_{obs}$, the data are effectively two-dimensional, as both $p$ and $t$ are important to the observed dynamics.
However, when $\lambda$ and/or $t_{obs}$ is large, the system rapidly equilibrates and the data are effectively one-dimensional, parametrized by time.
For this specific example, these changes are consistent with the analytically-predicted  shift in the dynamical behavior from the telegrapher equation to the diffusion equation.
From \eqref{eq:second_order_pde}, we can see that $t_{obs}$, the time scale of observation, should be larger than the time scale $1/\lambda$ for the diffusion equation limit in \eqref{eq:diff_eqn} to hold; we therefore expect a transition in the dynamical behavior at
\begin{equation}
t_{obs} \approx \frac{1}{\lambda}.
\end{equation}
This curve is also plotted in Figure~\ref{fig:chemotaxis_compare_timescales_evals} and is consistent with the transition predicted by the diffusion maps eigenvalues.

\section{Conclusions}

This paper addresses repeated eigendirections, a critical issue in applying diffusion maps to the analysis of complex dynamic data.
We show that, using local linear regression, we can automate the detection of eigenvectors corresponding to unique directions in the
geometry of the data.
From this detection, we can then obtain a {\em reduced diffusion maps embedding} which is both parsimonious and 
induces a metric which is equivalent to the standard diffusion distance.
We showed that this algorithm enables us to more fruitfully analyze data from a complex dynamical system, 
enabling the extraction of good reduced coordinates and the detection of changes in the dimensionality of the macroscopic dynamics.
We are confident that the proposed methodology will be helpful in the analysis of high-dimensional data sets for which 
the dimensionality of the underlying manifold is unknown.

\section*{Acknowledgments}
C.J.D. was supported by the Department of Energy Computational Science Graduate Fellowship (CSGF), grant number DE-FG02-97ER25308, and the National Science Foundation Graduate Research Fellowship, grant number DGE 1148900. 
R.T. was supported by the European Union’s Seventh Framework Programme (FP7) under Marie Curie Grant 630657 and by the Horev Fellowship.
I.G.K. was supported by the National Science Foundation (CS\&E program).

\bibliographystyle{elsarticle-num}
\bibliography{../../../references/references}

\end{document}